\documentstyle[11pt,twoside,fancyhea,emlines]{article}
\title{\Large\bf
INVESTIGATION OF THE ISING-TYPE MODELS WITHIN CLUSTER APPROACH
   }
\author{
   {\sc R.~R.~Levitskii, S.~I.~Sorokov}
   \\[1.5ex]
   \it Institute for Condensed Matter Physics
   \\
   \it of the Ukrainian Academy of Sciences
   \\
   \it 1~Svientsitski St., UA--290011 Lviv, Ukraine
}
\date{Received June 14, 1994}

\newcommand{\be}{\begin{equation}}
\newcommand{\ee}{\end{equation}}

\newcommand{\bee}{\begin{eqnarray}}
\newcommand{\eee}{\end{eqnarray}}
\newcommand{\ang}[1]{\langle #1\rangle}
\newcommand{\dif}[2]{\frac{\partial #1}{\partial #2}}
\newcommand{\fdif}[2]{\frac{\delta #1}{\delta #2}}
\renewcommand{\kappa}{\zeta}
\newcommand{\lav}[1]{\,^L\ang{#1}}
\newcommand{\set}[1]{{\bf #1}}
\newcommand{\myhbox}[1]{#1}
\renewcommand{\theequation}{\arabic{section}.\arabic{equation}}
\newcounter{xxx}

\newcommand{\beea}
{
\setcounter{xxx}{0}
\renewcommand{\theequation}{\arabic{section}.\arabic{equation}\addtocounter{xxx}{1}\alph{xxx}\addtocounter{equation}{-1}}
\bee
}

\newcommand{\eeea}{\eee
\renewcommand{\theequation}{\arabic{section}.\arabic{equation}}
\refstepcounter{equation}
}

\begin{document}
\setcounter{page}{59}
\maketitle

\begin{abstract}
{\small The method for calculation of the
correlation functions of the Ising-type systems with short-range
interaction  and with arbitrary value of spin is developed
within cluster approximation.
       For the Ising model (spin $S^z=\pm1$) the expressions for pair and
ternary correlation functions within two-particle approximation
 in $\set{q}$-space are obtained for the
hypercubic Bravais lattices. In the 1D case the exact expressions for them
in the site space is obtained as well. On the basis of the Glauber
equation within two-particle cluster approximation the longitudinal
dynamical susceptibility $\chi(\set{q},E)$ is found. In the 1D case and in
the absence of external field the expression for $\chi(\set{q},E)$ is
exact. For the Emery-Blume-Griffiths model ($S^z=-2,0,2$) within
two-particle approximation the pair correlation functions
are calculated. The four-particle cluster
approximation is used for calculation of static  susceptibility
$\chi(\set{q})$ of $KD_2PO_4$ ferroelectrics.}
\end{abstract}

\section{Introduction.}

	The wide class of materials is known  which is described on the
basis of the pseudospin Hamiltonian. In particular these are magnets with
different values of spin $M$~[1-3] (in our notations $S^z=-M, -M+2, \dots,
M-2, M$), ferroelectrics with hydrogen bonds~[4,5] ($M=1$), multicomponent
alloys~[6,7] ($M+1$ corresponds to the number of components), lattice gas~[8]
($M+1$ is the number of atom states per site). In the works~[9-12] the
method of investigation of pseudospin system based on the introducing the
reference system (RS) was developed. This approach makes possible to take
into account adequately both the short-range and long-range interaction.
RS contains besides the low-dimensional short-range ($D=1,2$) part of the
Hamiltonian the terms taking into account the other types of interactions
in the molecular field approximation (MFA). The fluctuating
corrections to MFA for these types of interactions are taken into account on
the basis of expansion of the physical responses (free energy, correlation
functions, temperature Green functions) in terms of loop diagrams or in
terms of many-tails diagrams. These corrections contain the physical
responses of RS. The calculation of the latters is the separate problem
of the referense system approach.

        In the series of cases for the description of pseudospin system's
properties it is enough to take into account the long-range interaction in
MFA. Now the reference system problem is to be considered separately.
However, only a small number of low-dimensional problems may be
solved exactly.
%\newpage
The exact solutions are obtained mainly in the case of 1D quantum models or
1D and 2D Ising-type models (ITM) (the Hamiltonian of ITM contains only $S^z$
component of the spin). Among approximate methods of investigation the
cluster method is sufficiently effective~[13-31]. It is necessary
to distinguish the cluster expansion method (CEM) and the cluster variation
method (CVM). In the case of CEM the cluster expansion is constructed for the
free energy~[13] whereas in the case of CVM one carries out the cluster
expansion of entropy~[26]. It can be shown that in the first order of
cluster expansion both methods give equivalent results for all responces of
pseudospin system. This approximation we shall call the cluster approximation
(CA). Let us note that in the first works within the framework of CVM the
results were obtained on the basis of combinatoric approach~[25] whereas within
CEM the results was obtained on the basis of many-particle generalization of
MFA~[14].

        Unfortunately the cluster approach was used mainly for investigation
of thermodynamical properties of pseudospin systems. We know only several
papers [32-43] where the cluster approach was used for calculation of
reference system's correlation functions (CF) (see~[32-37]) and Green
functions~[38-43]. In the works~[32,33] the equation for pair CF
(Ornstein-Zernike type equation) of ITM was constructed artificially. The
dipole-dipole interaction which has the nonanalytical properties at
$\set{q}\rightarrow 0$ was taken into account in MFA. The concrete
calculations were carried out only in the case $\set{q}\rightarrow 0$. Let us
note that only in paraphase the obtained results correspond to cluster
approximation. In the work~[34] within the frames of CA for ITM the method
for calculation of pair CF for RS, when the long-range interaction is absent,
was suggested. This method is applicable in the case $T>T_c$. Here analytic
and numerical investigation of $q$-dependence for CF of some models was
performed. The Ising model (IM) (IM corresponds to ITM with $M=1$) on square
lattice within the frame of two-particle cluster approximation (TPCA) and
four-particle cluster approximation (FPCA) was considered. The model of
ferroelectrics $KD_2PO_4$ was studied too. In the paper~[35] for IM
within the frame of cluster expansion method the method which allows to find
the correlation function of arbitrary order for RS within cluster
approximation was suggested. For IM on hypercubic lattices within TPCA
the pair and ternary CF were found. Here it was shown that for 1D system
these results coincide with the exact ones. In the ref.~[36] it was proved
that TPCA for IM yields exact results for all characteristics of the system
in the case of tree-like lattice. Moreover in the case of square lattice within
CEM the influence of the higher order terms of cluster expansion
on thermodynamic characteristics was studied numerically. The pair CF in FPCA
was also calculated. In the paper [37] CEM was generalized on the Ising type
systems with arbitrary value of spin and detailed numerical investigation of
the Blume-Emery-Griffits model was performed.

    In the works~[38-41] the dynamics of ferroelectrics with hydrogen bonds
with accounting of tunneling was investigated within the frames of cluster
approach. Here the method of two-time  Green's function was used. The equation
for them was uncoupled in the spirit of Tyablikov approximation with respect
to long-range interaction. At such consideration the short-range intercluster
correlations were not taken into account and intracluster Green's functions
appear to be connected only through long-range interaction. This leads to
uncorrespondence of expressions for static susceptibility which follows
(at $\set{q}=0$ and $\omega_n=0$) from the expressions for dynamic susceptibility with
expressions which are obtained within CA. In the works~[42,43]
a method was developed which within the frames of cluster approach allowed to obtain
selfconsistent results for thermodynamic characteristics and temperature Green's
functions. Within TPCA the reference Green's functions were calculated for the
Heisenberg model~[42] and the Ising model in a transverse field~[43].

     In the present work on the basis of cluster approximation the pseudospin
systems which are described by Ising type models, are investigated. In
section~2 the statement of CEM for ITM with arbitrary many-particle short-range
interaction is presented. In  section~3 within TPCA the thermodynamic,
correlation functions and relaxational dynamics of IM (M=1) on hypercubic
lattice are investigated. In section~4 in the frames of TPCA the Ising type
model with pair interaction and arbitrary value of spin is considered.
Section~5 is devoted to investigation within TPCA of the Ising type model
describing the ferroelectric properties of crystal $KD_2PO_4$.

\setcounter{equation}{0}
\section{Statement of the cluster expansion me\-thod.}

In this section we shall consider the cluster expansion method for the
Ising-type model with arbitrary form of the Hamiltonian. In the following
sections we shall apply present method to certain models. The Hamiltonian of
IM with arbitrary values of spin can be written in the following form
($S^z_i = S_i = - M, - M +2, \ldots, M - 2, M$)
\beea
&&- \beta \,^L{\cal H} = \,^LH(\{h^{(\cdot )}\}) = H(\{h^{(\cdot )}\}) + {1\over 2} \sum_{i,j} J_{ij}(S_i,S_j) ,\\
&&H(\{h^{(\cdot )}\}) = \sum_i h_i(S_i) +\sum_{(ij)} K_{ij}(S_i,S_j) + W(\{S\}) ,\\
&&h_i(S_i) = \sum^M_{\mu =1} h^{(\mu )}_i S^{\mu }_i ;\; K_{ij}(S_i,S_j) = \sum^M_{\mu =1} \sum^M_{\nu =1} K^{(\mu,\nu)}_{ij} S^{\mu }_i S^{\nu }_j \,,\\
&&J_{ij}(S_iS_j) = \sum^M_{\mu =1} \sum^M_{\nu =1} J^{(\mu ,\nu )}_{ij} S^{\mu }_i S^{\nu }_j \,.
\eeea
The one-site part of the Hamiltonian $h_i(S_i)$ describes interaction of
pseudospins with the fields of different type. The two-site part of the
Hamiltonian contains the short-range $K_{ij}(S_iS_j)$ and long-range
$J_{ij}(S_iS_j)$ pair interactions. The three-site, four-site etc. short-range
interactions are included in $W(\{S\})$. The pseudospin system with the
Hamiltonian $H(\{h^{(.)}\})$ is called the reference system~[9-12]. We shall
consider the RS which contains only the interaction with the nearest
neighbours. In particular, for pair interaction we can write
\be
K^{(\mu ,\nu )}_{ij} = K^{(\mu ,\nu )} \pi _{ij} ;\,\pi_{ij} = \left\{
\begin{array}{ll}
        1, & i $ is the nearest neighbour of $ j ,\\
        0, & $in opposite case.$
\end{array}
             \right.
\ee
\newcommand{\F}{{\cal F}}
In the case of the Bravais lattice the indices $i,j$ denote the sites of the
lattice. The case of the structure with sublattices we shall consider in
section 5. In the present paper the pair long-range interaction is taken
into account in the molecular field approximation (MFA). Now the $\F$-function
(the logarithm of partition function) of the system with the Hamiltonian
(2.1) can be written in the following form

\be
^L\F(\{h^{(\cdot )}\}) = - {1\over 2} \sum_{\mu  \nu } \sum_{i,j} J^{(\mu ,\nu )}_{ij} \,^L\ang{S^\mu_i} \,^L\ang{S^\nu_j} + \F(\{\kappa ^{(\cdot )}\}).
\ee
Here $\F(\{\kappa^{(\cdot)}\})$ is $\F$-function of RS with the Hamiltonian
containing the long-range molecular field $\lambda_i^{(\mu)}$:
\beea
&\F(\{\kappa ^{(\cdot )}\})= \ln  Z(\{\kappa ^{(\cdot )}\});\; Z(\{\kappa ^{(\cdot )}\})= Sp_{\{s\}} \exp  [H(\{\kappa ^{(\cdot )}\})]  ;\\
&\kappa ^{(\mu )}_i = h^{(\mu )}_i + \lambda ^{(\mu )}_i ;\qquad \lambda ^{(\mu )}_i = \sum_{j,\nu } J^{(\mu ,\nu )}_{ij} \,^L\ang{S^{\nu }_j}.
\eeea
Here the average $\lav{S_i^\nu}$ is taken for canonical ensemble and the
density matrix is constructed with the Hamiltonian (2.1a). We shall calculate
correlation functions (CFs) of RS (cumulant averages of products of operators
$S_i^{\mu}$) in the following way:
\bee
\F^{(l)}\left(^{\mu_1..\mu_l}_{\i_1..i_l};\{\kappa ^{(\cdot )}\}\right)
= \ang{S^{\mu _1}_{i_1} \ldots S^{\mu_l}_{i_l} }^c = \nonumber\\
= {\delta \over \delta  \kappa ^{(\mu _1)}_{i_1}} \ldots
 {\delta \over \delta  \kappa ^{(\mu _l)}_{i_l}} \F(\{\kappa ^{(\cdot )}\}).
\eee
Similarly to (2.5) we shall obtain CFs
$^L\F^{(l)}(^{\mu_1..\mu_l}_{i_1..i_l};\{h^{(\cdot)}\})$ for system with long-range
interaction. In this case the $\F$-function $^L\F(\{h^{(\cdot)}\})$ is
differentiated with respect to fields $h^{(\mu)}_j$. In the MFA these CFs can
be connected with the CFs of RS. In particular for $\lav{S_i^{\mu}}$ using the
relations (2.3), (2.5) we obtain the equality
\be
^L\F^{(1)}\left(^\mu_i;\{h^{(\cdot )}\} \right)
= \,^L\ang{S^{\mu }_i} = \ang{S^{\mu }_i} = \F^{(1)}\left(^\mu_i;\{\kappa^{(\cdot )}\} \right).
\ee
If we shall differentiate (2.6) and take into account the functional
dependence of $\kappa^{(\mu)}$ on $\ang{S_{i'}^{\mu'}}$, we shall obtain the
relation for pair CFs. It has the following matrix form
\be
^L\hat{\F}^{(2)}(\{h^{(\cdot )}\}) = \hat{\F}^{(2)}(\{\kappa ^{(\cdot )}\}) \left[\matrix{1-\hat{J}\hat{\F}^{(2)}(\{\kappa ^{(\cdot )}\})}\right]^{-1}\,.
\ee
Here the matrices $\hat{A}=\{A_{ij}^{(\mu,\nu)}\}$ are constructed with the help
of power-type indices $\mu$, $\nu$ and indices of sites $i$, $j$. In the
homogeneous field $h_i^{(\mu)}=h^{(\mu)}$ the Fourier transformation
diagonalizes the relation (2.7) with respect to the indices of the elementary cell.

        At first let us decompose lattice into clusters for construction of
cluster expansion of $F$-function~[36]. The figure~1 demonstrates decomposition of
square lattice into clusters.
\begin{figure}
\begin{center}	\input fig1.pic	\end{center}
\caption[]{
The decomposition of square lattice into clusters ("link", "square")
and demonstration of effects of cluster fields on site. Here "s" is the
number of clusters containing present site, "b" is the number of clusters
containing present link, "k" is the number of sites in cluster.%\\[1ex]
}
\end{figure}

	Let us note that certain choice of lattice decomposition
depends on the number of factors. In particular, this choice depends on
structure and symmetry of crystal, on form of interaction and cluster
expansion order. In the present work we shall use only lattice decomposition
into two-particle ("link") and four-particle (tetrahedron) clusters. We shall
denote the effective field operator as $_R\varphi_i(S_i)$. It effects on
site $i$ from the side of cluster $R$, which contains this site. The set of
these clusters we shall mark $\pi_i$. Obviously, $s_i$ fields effect on
an arbitrary site $i$. After decomposition of lattice into clusters we
pass from summation over sites to summation over clusters.
\bee
&&\sum_i \sum_R \,_R\varphi _i(S_i) = \sum_R \sum_{f\in R} \,_R\varphi _f(S_f); \;W(\{S\}) = \sum_R W_R(\{S\})\,, \nonumber\\
&&\sum_{(ij)} K_{ij}(S_iS_j) = \sum_R \sum_{\myhbox{ff'}\in R} {1\over b_{\myhbox{ff'}}} K_{\myhbox{ff'}}(S_fS_{f'}).
\eee
Here we shall consider the cases when $W_R(\{S\})$ contains only spins from
the cluster $R$. Taking into account (2.8) after identical transformation the
Hamiltonian (2.1) is represented in the following way
\be
H = \sum_i H_i(S_i) + \sum_R U_R(\{S\})\,,
\ee
where
\bee
&&H_i(S_i) = h_i(S_i) + \sum_R \,_R\varphi _f(S_f)\,, \nonumber\\
&&U_R(\{S\}) = - \sum_{f\in R} \,_R\varphi _f(S_f) + \\
&&+\sum_{\myhbox{ff'}} {1\over b_{\myhbox{ff'}}} K_{\myhbox{ff'}}(S_fS_{f'}) + W_R(\{S\}).\nonumber
\eee
Then, using the form (2.9) and cluster expansion method (CEM)~[13], the
$\F$-function can be written in the following form
\be
\F = \sum_i F_i + \ln  \ang{\exp \left(\sum_R U_R(\{S\})\right)}_0
= \sum_i F_i + \sum^{\infty }_{l=1} K_l \,.
\ee
Here we use the notations
\be
F_i = \ln  Z_i = \ln  Sp e^{H_i} ;\qquad \rho _0 =\prod_i \rho_i =\prod_i {e^{H_i}\over Z_i}\,.
\ee
Functions $K_\set{l}$ contain sums over sets $\set{l}=(R_1\cdots R_l)$,
which include noncoinsiding clusters $R_i$.
\bee
&&K_l =\sum_{(R_1,\ldots,R_l)} K(R_1,\ldots
,R_l) = \sum_{\{{\bf l\}}} K({\bf l})\,, \nonumber\\
&&K({\bf l}) = \sum^{\infty }_{\nu _1=1} \ldots
 \sum^{\infty }_{\nu _m=1} {1\over \nu _1!} \ldots
 {1\over \nu _m!} \ang{U^{\nu _1}_{R_1}\ldots U^{\nu _m}_{R_m}}_0 \,.
\eee
The cluster functions $K(\set{l})$, as it is well known, are expressed by
means of $L(\set{m})$-functions as well as functions $L(\set{l})$ are
expressed by means of $K(\set{m})$ in the following way~[13]
\beea
K({\bf l})= \sum^l_{m=1}(-1)^{l-m} \sum_{{\bf m\subset l}} L({\bf m}) = L(l) + \sum^{l-1}_{m=1}(-1)^{l-m} \sum_{{\bf m\subset l}} L({\bf m}) \,,\\
L({\bf l})= \sum^l_{m=1} \sum_{{\bf m\subset l}} K({\bf l}) = K({\bf m}) + \sum^{l-1}_{m=1} \sum_{{\bf m\subset l}} K({\bf m}) \,,
\eeea
where $L$-function has the form ($M(\set{l})$-moment function):
\bee
&&L({\bf l}) = \ln  M({\bf l}) = F({\bf l}) - \sum_{i \in \sum^l_{n=1}R_n} F_i \,, \nonumber\\
&&M({\bf l}) = \ang{\exp  \left[ \sum^l_{n=1} U_{R_n}\right]}_0 \,.
\eee
Here we use the notations $\set{l}$ for set of clusters, $F(\set{l})$ for
$\F$-function, $Z(\set{l})$ for partition function and $H(\set{l})$ for the
Hamiltonian:
\bee
&&F({\bf l}) = \ln  Z({\bf l}) = Sp_{\{S\}} \exp  \{H({\bf l})\} \,, \nonumber\\
&&H({\bf l}) = \sum_{i \in \sum^l_{n=1}R_n} H_i + \sum^l_{n=1} U_{R_n} \,.
\eee
Arbitrary set of clusters $\set{l}$ of present type forms certain graph
(diagram)~[36] on the lattice. We shall call graph to be unconnected one if it has
parts which do not contain common sites. These parts will be independent
statistically, when we shall carry out the averaging with distribution
function $\rho_0$. Let us separate arbitrary graph $\set{n}$ into two
graphs $\set{n}_1$ and $\set{n}_2$ which have not comon clusters, but which
may have common sites.
\unitlength=1.00mm
\special{em:linewidth 0.4pt}
\linethickness{0.4pt}
\be
\raisebox{-3mm}{
\begin{picture}(50,10)
\put(-27,10){\special{em:graph 217a.pcx}}
\put(-20,3){\makebox(0,0)[lc]{${\bf n}_1\qquad\;\;\;{\bf n}_2
\qquad\;\;\; {\bf n} = {\bf n}_1 \cup  {\bf n}_2,
\qquad {\bf n}_1 \cap  {\bf n}_2 = \set{M}_{12}$} }  .
\end{picture}    }
\ee
 \\[2ex]
Here $\set{M}_{12}$ is the set of sites, which is common for $\set{n}_1$
and $\set{n}_2$. Let us represent the sum $\sum_{\set{m}\subset \set{n}}$ in
the form $\sum_{\set{m}_1}\sum_{\set{m}_2}$, where $\set{m}_1\subset\set{n}_1$,
$\set{m}_2\subset\set{n}_2$. Then we can write (after separation of items with
$\set{m}_1=0$ and $\set{m}_2=0$ \,)
\be
K({\bf n})=L({\bf n})-L({\bf n}_1)-L({\bf n}_2)
- \sum^{n_1}_{m_1=1} \sum^{n_2}_{m_2=1} \sum_{{\bf m}_1\subset {\bf n}_1}
\sum_{{\bf m}_2\subset {\bf n}_2} K({\bf m}_1+{\bf m}_2) \,.
\ee
The prime near the symbol of sum means that item at
$m_1=n_1$, $m_2=n_2$ is omitted. Let the following condition take place
\be
L({\bf m}_1 + {\bf m}_2) = L({\bf m}_1) + L({\bf m}_2),\qquad \forall  {\bf m}_1 \subset  {\bf n}_1\,, {\bf m}_2 \subset  {\bf n}_2.
\ee
Then we can carry out recurrent procedure for $K(\set{m}_1+\set{m}_2)$,
which begins with $m_1=1$, $m_2=1$. For
$\set{m}_1=\set{2}_1=\set{1}_1+\set{1'}_1$, $\set{m}_2=\set{1}_1$ we have
\bee
K({\bf 1}_1+{\bf 1}_2) &=& L({\bf 1}_1+{\bf 1}_2) - L({\bf 1}_1) - L({\bf 1}_2) \equiv  0\,, \nonumber\\
K({\bf 1}_1'+{\bf 1}_2) &=& L({\bf 1}_1'+{\bf 1}_2) - L({\bf 1}_1' ) - L({\bf 1}_2) \equiv  0\,, \\
K({\bf 2}_1+{\bf 1}_2) &=& L({\bf 2}_1+{\bf 1}_2) - L({\bf 2}_1) - L({\bf 1}_2) \nonumber\\
&&- K({\bf 1}_1+{\bf 1}_2) - K({\bf 1}_1'+{\bf 1}_2)\equiv 0. \nonumber
\eee
Similarly we can obtain
\be
K({\bf m}_1 + {\bf m}_2) \equiv  0,\qquad \forall  {\bf m}_1 \subset  {\bf n}_1\,,\;\forall {\bf m}_2 \subset  {\bf n}_2 \,.
\ee
The condition (2.19) is fulfiled for $\set{n}_1$ and $\set{n}_2$ being two
unlinked parts of graph $\set{n}$ (when $\set{M}_{12}=0$). The condition
can be fulfiled also for some linked graphs, if we choose the fields
$_R\varphi_i$ by certain manner (see the following section or paper~[36]).
Let's note~[36] that for set (2.14a) some $L$-functions, which
correspond to unlinked diagrams are eliminated by $L$-functions of linked
diagrams of certain type. Therefore there is a sum over linked graphs of
certain class in (2.14a).

        In the present paper we shall use mainly the first approximation
for $F$-function and the decomposition of the lattice into R-particle
clusters (RPCA). As it follow from (2.11)--(2.16) the $F$-function in RCPA
has the form~[36]
\be
\F(\{\kappa ^{(\cdot )}\}) = - \sum_i (s_i - 1) F_i(\{\tilde{\kappa }^{(\cdot )}\})
+ \sum_R F_R(\{_{R'}\tilde{\kappa }^{(\cdot )}\}) \,.
\ee
Here the notations for intracluster one-particle\,$(F_i)$\, and R-particle \,$(F_R)$\,
$\cal F$-functions and corresponding partition function $Z_i$, $Z_R$ are
introduced
\beea
&&F_i(\{\kappa ^{(\cdot )}_i\}) = \ln  Z_i(\{\tilde{\kappa }^{(\cdot )}_i\}) ;\; Z_i(\{\tilde{\kappa }^{(\cdot )}_i\}) = Sp_{S_i} e^{H_i(S_i)} ;\\
&&F_R(\{_{R'}\kappa ^{(\cdot )}\})=\ln  Z_R(\{_{R'}\tilde{\kappa }^{(\cdot )}\}) ;\; Z_R(\{_{R'}\tilde{\kappa }^{(\cdot )}\})=Sp_{\{S\}} e^{H_R(S)}\;.
\eeea
The one-particle Hamiltonian $H_i(S_i)$ and R-particle Hamiltonian
$H_R(\{S\})$ have the form
\beea
&H_i(S_i) = \sum_{\mu } \tilde{\kappa}^{(\mu )}_i S^{\mu }_i ;\qquad \tilde{\kappa }^{(\mu )}_i = \kappa ^{(\mu )}_i + \sum_{R'} \,_R'\varphi ^{(\mu )}_i ;\\
&H_R(\{S\}) = \sum_{f\in R} H_j(S_j) + U_R(\{S\})=  \nonumber\\
&= \sum_{f\in R} \,_R\tilde{\kappa }_f(S_f) + \sum_{\myhbox{ff'}\in R} {1\over b_{\myhbox{ff'}}} V_{\myhbox{ff'}}(S_fS_{f'}) + W_R(\{S\}) ;\\
&_R\tilde{\kappa }_f(S_i) = \sum_{\mu } \;_R\tilde{\kappa }^{(\mu )}_f S^{\mu }_f \;; \,_R\tilde{\kappa }^{(\mu )}_f = \kappa ^{(\mu )}_f +\sum_{R'\neq R} \,_{R'}\varphi ^{(\mu )}_f.
\eeea
As we can see from (2.24c), the fields $_R\varphi_f^{(\mu)}$ which effect
on site $f\subset \set{R}$ from cluster $\set{R}$ are not contained in
R-particle Hamiltonian. This is naturally due to the fact that all
these interactions are taken into account exactly with the help of terms
$V_{ff'}$ and $W_R$. On the basis of Hamiltonians $H_i$, $H_R$ we can
introduce intracluster density matrices
\be
\rho _i(S_i) = {e^{H_i(S_i)} \over Z_i}; \; \rho _R(\{S\}) = {e^{H_R(\{S\})} \over Z_R}\,,
\ee
and intracluster CFs
\beea
&&\ang{(S_i)^{l^i_1} (S^2_i)^{l^i_2} \ldots (S^M_i)^{l^i_M}}^c_{\rho _i}\,, \\
&&\ang{(S_{f_1})^{K^{f_1}_1} \ldots (S^M_{f_1})^{K^{f_1}_M} \ldots
 (S_{f_R})^{K^{f_R}_1} \ldots (S^M_{f_R})^{K^{f_R}_M} }_{\rho _R}^c \,.
\eeea
We shall calculate CFs (2.26) on the basis of $F$-functions (2.23) and
relations of (2.4)-type. Evidently, they do not equal zero only if the site
indices ($i$ or $f$) belong to present cluster ($\set{i}$ or $\set{R}$).

        We shall find the system of equations for the fields $_R\varphi_i^{(\mu)}$
from condition of stationarity of $\cal F$-function with respect to
these fields. In the present paper we shall consider only RPCA. As it will
be written in the following sections the system of $Ns_iM$
equations in RCPA  has the form
\be
\ang{S^\mu_i}_{\rho_i} = \ang{S^\mu _i}_{\rho _R}\,.
\ee
Here the average is defined with respect to density matrices (2.25). The
system (2.27) is equivalent to the system of $Ns_i(M+1)$ equations
\be
\rho _i(S_i) = Sp_{{\bf R-i}}\, \rho _R(\{S\})\,,
\ee
which coincides with the exact relations between one particle and
R-particle density matrices. Among $Ns_i(M+1)$ equations (2.28) only
$Ns_iM$ ones are independent because $Ns_i$ conditions
$Sp_iSp_{{\bf R-i}}\,\rho_r(\{S\})=1$ are fulfiled identically. In the
homogeneous
field ($h_i=h$, $\kappa_i=\kappa$, $_R\varphi_i^{(\mu)}=\varphi^{(\mu)}$)
(2.27) gives $M$ equations for $M$ unknown fields $\varphi^{(\mu)}$.

\setcounter{equation}{0}
\section{Thermodynamic \ and \ dynamic \ properties of the Ising model.
Two-particle cluster ap\-pro\-xi\-ma\-ti\-on}
\subsection{The main relations. Thermodynamic properties}

In the present section we shall consider the Ising-type model at $M=1$,
$S_i=\pm 1$ (Ising model-IM) in the case of hypercubic lattice. Now the
Hamiltonian (2.1) contains only linear and bilinear forms of pseudospin
operators and we can write it in the following way
\be
H(\{\kappa \}) = \sum_i \kappa _iS_i +\sum_{(ij)} K_{ij} S_iS_j ;\qquad K_{ij} = \beta K\pi _{ij}\,.
\ee
In TPCA the cluster contains only two sites ($\set{R}_{1r}=(1,r)$) and we
can speak about effect on the site $i$ from it's neighbour $r$. This fact
leads to the change of notations for the cluster fields:
$_R\varphi_1\rightarrow\,_r\varphi_1$. Thus, taking into account
(2.22)--(2.24), we obtain ($s_i=z$, $b_{ff'}=1$)~[35]
\be
\F(\{\kappa \}) = (1-z) \sum_1 F_1(\tilde{\kappa }_1) + {1\over 2} \sum_{1,r} F_{1r}(_r\tilde{\kappa }_1\mid _1\tilde{\kappa }_r).
\ee
The one-particle and two-particle cluster Hamiltonians have the form
\beea
&H_1(S_1) = \tilde{\kappa }_1 S_1 ;\; \tilde{\kappa }_1 = \kappa _1 + \sum_{r'\in \pi _1} \,_{r'}\varphi _1\,, \nonumber\\
&H_{12}(S_1,S_2) = _2\tilde{\kappa }_1 S_1 + _1\tilde{\kappa }_2 S_2 + K S_1S_2\,, \\
&_2\tilde{\kappa }_1 = \kappa _1 + \sum_{r'\in\pi_1;\;r'\ne 2} \,_{r'}\varphi _1 ;\; _1\tilde{\kappa }_2 = \kappa _2 + \sum_{r'\in\pi_2;r'\ne 1} \,_{r'}\varphi _2.
\eeea
Thus, the following expressions are obtained for the intracluster
$\F$-func\-ti\-ons:
\beea
&F_1(\tilde{\kappa }_1) = F^{(0)}_1 = \ln  \{2\cosh \tilde{\kappa }_1\};\\
&F_{1r}= F^{(00)}_{1r}= \nonumber\\
&=\ln \left\{2e^{\beta K} [\cosh(_r\tilde{\kappa }_1 + \,_1\tilde{\kappa }_r) + e^{-\beta K} \cosh(_r\tilde{\kappa }_1 - \,_1\tilde{\kappa }_r)]\right\}\,.
\eeea
If we take into account the long range interaction in MFA we must
renormalize the field $\kappa_i$:
$\kappa_i=h_i+\lambda_i=h_i+\sum_jJ_{ij}\ang{S_j}$. Here $h_i$ is an
external field, $\lambda_i$ a long-range-acting molecular field.

        In the present section we shall use the following notations
\beea
&&\F^{(l)}_{i_1 \ldots i_R}(\{\kappa \}) = \ang{S_{i_1}\ldots
S_{i_R}}^c = {\delta \over \delta \kappa _{i_1}} \ldots
 {\delta \over \delta \kappa _{i_l}}\F(\{\kappa \})\,, \\
&&\tilde{\kappa}^{(1)}_{ij} = \delta _{ij} + \sum_{r\in \pi _i} \,_r\varphi ^{(1)}_{ij} = {\delta \over \delta \kappa _j} \tilde{\kappa }_i, \nonumber\\
&&\tilde{\kappa }^{(l-1)}_{i_1i_2\ldots
i_l} = {\delta \over \delta \kappa _{i_2}} \ldots
 {\delta \over \delta \kappa _{i_l}} \tilde{\kappa }_{i_1}= \sum_{r\in \pi _{i_1}} \,_r\varphi ^{(l-1)}_{i_1\ldots
i_l} ;\; l\ge 3 \,,\\
&&_r\varphi ^{(l-1)}_{i_1\ldots
i_l} = {\delta \over \delta \kappa _{i_2}} \ldots
 {\delta \over \delta \kappa _{i_l}} \,_r\varphi_{i_1} \,,\nonumber\\
&&F^{(l)}_{i_1}(\tilde{\kappa }_1)
= \dif{^l_1}{(\tilde{\kappa }_1)^l} F^{(0)}_1(\tilde{\kappa }_1) = \ang{S^l}^c_{\rho _1}\,, \nonumber\\
&&F^{(l_1l_2)}_{12}(_2\tilde{\kappa }_1\mid _1\tilde{\kappa }_2) =
\dif{^{l_1}}{(_2\tilde{\kappa }_1)^{l_1}} \dif{^{l_2}}{(_1\tilde{\kappa }_2)^{l_2}}  F^{(00)}_{12} =
\ang{S^{l_1}_1 S^{l_2}_1}^c_{\rho _{12}} \,.
\eeea
Now taking into account (2.5) we easily obtain from (3.2)
\be
\ang{S_1} = \F^{(1)}_1(\{\kappa \}) = \frac{\partial}{\partial\kappa _1}\F(\{\kappa \}) + \sum_{1',r} {\partial \over \partial_r\varphi _{1'}} \F(\{\kappa \}) \,_r\varphi ^{(1)}_{1'1}\,.
\ee
\vspace{-0.3em}
Here
\vspace{-0.5em}
\beea
&&\dif{}{\kappa _1}\F(\{\kappa \}) = (1-z) F^{(1)}_1 + \sum_{r'} F^{10}_{1r'}\, ,\\
&&\dif{}{_r\varphi _{1'}}\F(\{\kappa \})
= (1-z) F^{(1)}_{1'} + \sum_{r'} F^{10}_{1'r'} - F^{10}_{1'r}\, .
\eeea
Proceeding from stationarity condition for $\F(\{\kappa\})$ with respect to
$_r\varphi_1$ ($\dif{}{_r\varphi_1}\F(\{\kappa\})=0$) on the basis of (3.6)
and (3.7) we find the following relations
\beea
&m_1 = \ang{S_1} = \F^{(1)}_1(\{\kappa \}) = F^{(1)}_1(\{\tilde{\kappa }_1\}) \, ,\\
&\ang{S_1}_{\rho _1} = F^{(1)}_1(\{\tilde{\kappa }_1\})
= F^{(10)}_{1r}(_r\tilde{\kappa }_1\mid _1\tilde{\kappa }_r) = \ang{S_1}_{\rho _{1r}}\, .
\eeea
If the long-range interaction is absent (3.8b) gives the system of $Nz$
equations for $Nz$ unknown variational fields $_r\varphi_i(\{\kappa\},K)$
while (3.8a) gives the expression for the average value $\ang{S_1}$. The
expression (3.8b) is equivalent to the relation between one-particle $\rho_1$
and two-particle $\rho_{1r}$ density matrices:
\be
\rho _1(S_1) = Sp_{S_r} \rho _{1r}(S_1,S_r).
\ee
This equivalence can be proved if one takes into account that among
$2Nz$ equations (3.9) only $Nz$ equations are independent because $NZ$
conditions
$Sp_{S_1S_r}\rho_{1r}(S_1,S_r)$ $=Sp_{S_1}\rho_1(S_1)=1$ are
accomplished. It follows from (3.9) that
\be
F^{(l)}_1(\tilde{\kappa }_1) = \ang{(S_1)^l}^c_{\rho _1} = \ang{(S_1)^l}^c_{\rho _{12}} =
F^{(l,0)}_{1r} (_r\tilde{\kappa }_1\mid _1\tilde{\kappa }_r)
\ee
The equation (3.9) can be written in the other form which is useful for
diagram analysis
\be
M_{(1r)}= Sp_{S_1S_r}(\rho_1\rho_r e^{U_{1r}}) = Sp_{S_r}\rho_r e^{U_{1r}}\,.
\ee
Let us consider in expansion for $\F$-function certain type of graphs which
contains links that connect the site 1 with another part of the present
graph~[36]. In particular, this type of diagrams includes tree-graph, that is
graph of links without loops.
\be
\raisebox{-7mm}{
\begin{picture}(50,15)
\put(-23,15){\special{em:graph 312a.pcx}}
\put(25,15){\special{em:graph 312b.pcx}}
\end{picture}    
}
\ee

Let us denote by $\set{\nu}$ the set of all sites of graph $\set{n}$. Then for the
moment function of graph (3.12) taking into account (3.11) we can write
\be
M({\bf n}+(1r))=Sp_{\{S_j\}}\prod_{j\in \nu}\rho_j \exp(\sum_{\set{R}\in \set{n}} U_R) Sp \rho _re^{U_{1r}} = M({\bf n}) M(1r).
\ee
Since $M$-function for the graph (3.12) is factorized, the corresponding
$L$-function has the additive form (2.19). This fact means that for the
graph of (3.12)-type the $K$ function is equal to zero. It is shown in~[36]
that the differentiation $\dif{}{_r\varphi_1}$ of $K$-function for tree type
graphs also gives zero. But for $K$-function of arbitrary graph this
condition is not fulfilled (when the index $i$ in operator
$\dif{}{_k\varphi_i}$ coincides with $r$ in (3.12) and $\set{n}$ does not
contain links with free tails). Thus if in the expansion of $\F$-function we
shall consider only tree-graph and function $_r\varphi_i(\{\kappa\},K)$
will satisfy the system (3.11) (or identical system (3.8b)) then the
stationarity of $\F$-function with respect to fields will have been
accomplished. In this case the expressions for $\F$-function and $\ang{S_1}$
coincide with expressions (3.2) and (3.8a) which are obtained in TPCA. It
follows that for the lattice of the tree type, e.g. for 1D chain, the TPCA
gives an exact results for all quantities of the IM with Hamiltonian (3.1)
[35,36].

        Now we pay attention to thermodynamical properties of IM in TPCA
taking into account the long-range interaction in MFA. In the
case of homogeneous field extracting $\beta$ we can write
\beea
&&_r\varphi _i = \beta \varphi  ;\, \sum_r \,_r\varphi _i = z \beta \varphi ;\, \kappa _i = \beta \kappa  = \beta h + \beta J_0 m;\\
&&\tilde{\kappa }_i = \tilde{\kappa } = \beta (\kappa  + z \varphi ); \;_r\tilde{\kappa }_i = \beta \tilde{\tilde{\kappa }}= \beta [\kappa  +(z-1) \varphi ];
\eeea
where $J_0=J(\set{q}\rightarrow\set{0})$. From (3.2), (2.23) and (3.14) for the free
energy per site $f(h,K)$ we obtain the following result
\bee
&\beta\, ^Lf(h,K,J) = {1\over 2} J_0m^2 + \beta  f(\kappa ,K)\,,\nonumber\\
&-\beta f(\kappa,K) = {1\over N}\F(\kappa ) = (1-z) \ln Z_1(\tilde{\kappa }) + {z\over 2} \ln Z_{12}(\tilde{\tilde{\kappa}}).
\eee
Here the following notation is used
\bee
&&Z_1(\tilde{\kappa }) = 2\cosh (\beta \tilde{\kappa }) ;\qquad a = e^{-2\beta K}\, ; \nonumber\\
&&Z_{12}(\tilde{\tilde{\kappa}}) =2 e^{\beta K} [\cosh (2\beta \tilde{\tilde{\kappa}}) +a] \, .
\eee
The system of equations (3.8) for $m$ and $\varphi$ has the form
\beea
&&m = \tanh [ \beta (\kappa  + z\varphi )] ;\; z\varphi  = - \kappa  + (2\beta )^{-1} \ln  {1+m\over 1-m}; \\
&&\tanh[ \beta (\kappa  + z\varphi )] = \sinh (2\beta \tilde{\tilde{\kappa}}) [\cosh (2\beta \tilde{\tilde{\kappa}}) + a]^{-1}\, .
\eeea
In the case $J_0=0$ the system (3.17) leads to single equation for
$\varphi(\beta h, \beta K)$ (or for $m(\beta h, \beta K)$) if we exclude
$m$ (or $\varphi$) from (3.17a). In the 1D case ($J_o=0$, $z=2$) the equation (3.17b) for $\varphi$
can be solved exactly, and for $m$ and $f$ from (3.15) and (3.17a) we get
the generally known exact results~[44,45]
\beea
&&- \beta f = K + \ln  [\cosh \beta K + r] \,,\\
&&m = {\sinh(\beta h)\over r} ;\qquad r = \sqrt{a^2 + \sinh^2(\beta h)}\, .
\eeea
The equation (3.17b) can be used to obtain expressions for
$\dif{\varphi}{\kappa}$, $\dif{\varphi}{h}$. It makes possible to find the
pair correlation function $(\ang{S_iS_j}^c)_{\set{q}}=F^{(2)}_{\set{q}}$ at
$\set{q}=0$ and longitudinal isothermic susceptibility
$\chi^{zz}(\set{q}=0)=\chi^{zz}_0$ for IM:
\be
^L\F^{(2)}_{\set{q}=0} = {\delta m\over \delta h} = [ (\F_0^{(2)})^{-1} - \beta J_0]^{-1};\; \chi ^{zz}_0 = \beta {d^2\over \upsilon _a} \,^L\F^2_{\set{q}=0}\, .
\ee
Here $d$ is the dipole electric (or magnetic) moment per one cell, $v_a$
is the volume of elementary cell,
\be
(\F_0^{(2)})^{-1} = {z\over 2} {[a + \cosh(2\beta \tilde{\tilde{\kappa}})]^2\over 1 + a \cosh(2\beta \tilde{\tilde{\kappa}})} - {z - 1\over 1 - m^2}\, .
\ee
In the 1D case from (3.20) using solution for $\varphi$ one obtains the
expression
\be
(\F_0^{(2)})^{-1} = r^3 [a^2 \cosh(\beta h)]^{-1}\, .
\ee
On the basis of (3.15) one can easy obtain other thermodynamic quantities
of IM in TCPA, in particular, entropy and specific heat. In the case $h=0$
the equation (3.17) for $\varphi$ (or $m$) always has the solution
$\varphi=0$ ($m=0$) and at $T<T_c$ (below the phase transition temperature)
these equations have also non-trivial solution $\varphi\neq 0$ ($m\neq 0$).
As it was shown in~[36], in TCPA for IM at $K>0$ the phase transition (PT)
of the second order takes place. The equation for $T_c$ is found from
divergence condition for $^L\F_0^{(2)}$
\be
z \exp  [-2\beta _cK] - z + 2 = 2\beta _cJ_0 \, .
\ee
At $J_0=0$ from (3.22) one obtains well known result for $T_c$ of IM in
TCPA~[44,45]
\[
k_BT_c = 2K \left[\ln {z\over z-2}\right]^{-1} \, .
\]
In the 1D case ($z=2$) from (3.23) we find that $T_c=0$. Thus for 1D IM the
PT is absent at temperatures above zero.

\subsection{The correlation functions}

        In the present section we are going to consider briefly the
suggested method for calculation of the correlation functions of IM in the
cluster approach. We are going to find the explicit expressions for the pair
and ternary CFs in the wave vector space. The results and the method of
calculation of CFs in the site space in 1D case are presented in the
work~[35].

        On the basis of relations (2.5) and (3.8a) CFs can be expressed
with the help of derivatives of fields $\tilde{\kappa}_i$ (see (3.5b)) with
respect to $\kappa_j$. We present the general expressions for pair,
ternary and four-spin CFs in the wave vector space~[35]
\beea
&\F^{(2)}(\set{q}_1) = F^{(2)} \tilde{\kappa }^{(1)}(\set{q}_1)\,, \\
&\F^{(3)}(\set{q}_1\set{q}_2) = F^{(3)} \tilde{\kappa }^{(1)}(\set{q}_1) \tilde{\kappa }^{(1)}(\set{q}_2) + F^{(2)} \tilde{\kappa }^{(2)}(\set{q}_1\set{q}_2)\,, \\
&\F^{(4)}(\set{q}_1\set{q}_2\set{q}_3) = F^{(4)} \tilde{\kappa }^{(1)}(\set{q}_1) \tilde{\kappa }^{(1)}(\set{q}_2) \tilde{\kappa }^{(1)}(\set{q}_3) +   \nonumber\\
&+ F^{(2)} \tilde{\kappa }^{(3)}(\set{q}_1\set{q}_2\set{q}_3) + F^{(3)} \left [\tilde{\kappa }^{(1)}(\set{q}_1) \tilde{\kappa }^{(2)}(\set{q}_2\set{q}_3) +\right.   \nonumber\\
&\left. + \tilde{\kappa }^{(1)}(\set{q}_2) \tilde{\kappa }^{(2)}(\set{q}_1\set{q}_3) + \tilde{\kappa }^{(1)}(\set{q}_3) \tilde{\kappa }^{(2)}(\set{q}_1\set{q}_2)\right ] \,,\\
&F^{(2)} = 1-m^2; F^{(3)} = -2m F^{(2)}; F^{(4)} = - 2F^{(2)}(1-3m^2).
\eeea
Thus it is necessary to construct the equations for
$\tilde{\kappa}^{(1)}(\set{q})$,
$\tilde{\kappa}^{(2)}(\set{q_1},\set{q_2})$ and
$\tilde{\kappa}^{(3)}(\set{q_1},\set{q_2},\set{q_3})$ (on the basis of (3.8b))
for the computation of $\F^{(2)}(\set{q})$,
$\F^{(3)}(\set{q_1},\set{q_2})$ and $\F^{(4)}(\set{q_1},\set{q_2},\set{q_3})$.

        Now we are going to consider the equation for
$\tilde{\kappa}^{(1)}(\set{q})$. Differentiating (3.8b) with respect to
$\kappa_2$ one obtains
\beea
&&F^{(2)}_1 \tilde{\kappa }^{(1)}_{12} = F^{(20)}_{1r} \,_r\tilde{\kappa }^{(1)}_{12} + F^{(11)}_{1r} \,_1\tilde{\kappa }^{(1)}_{r2}\,, \\
&&_r\tilde{\kappa }^{(1)}_{12} = \tilde{\kappa }^{(1)}_{12} - \,_r\varphi ^{(1)}_{12};
\;_1\tilde{\kappa }^{(1)}_{r2} = \tilde{\kappa }^{(1)}_{r2} - \,_1\varphi ^{(1)}_{r2}\,.
\eeea
Taking into account (3.24b) and (3.10) the equation (3.24a) and the same
equation with replaced indices $1\leftrightarrow r$ constitute the system of
equations for the fields $_r\varphi_{12}^{(1)}$ and $_1\varphi_{r2}^{(1)}$.
This system can be written in the form
\be
\pmatrix{1&f_{1r}\cr f_{r1}&1} \pmatrix{_r\varphi_{12}^{(1)}\cr _1\varphi_{r2}^{(1)}} =
\pmatrix{f_{1r}\,\tilde{\kappa}^{(1)}_{r2}\cr f_{r1}\,\tilde{\kappa }^{(1)}_{12}} ; f_{1r} = {F^{(11)}_{1r}\over F^{(20)}_{1r}}
\ee
From (3.25) one can easy express $_r\varphi_{ij}^{(1)}$ in terms of
$\tilde{\kappa}_{ij}^{(1)}$. Taking into consideration the relationship (3.5b)
between $\kappa_{ij}^{(1)}$ and $\sum_r\,_r\varphi_{ij}^{(1)}$ one can
obtain the equation for $\kappa_{ij}^{(1)}$.
\be
\Theta _{11} \tilde{\kappa }^{(1)}_{12} = \delta _{12} +
\sum_r f_{1r}\, d^{-1}_{1r1} \,\tilde{\kappa }^{(1)}_{r2}\,.
\ee
Here the following notations are introduced
\be
\Theta _{11} = \sum_r {1+(z-1) f_{1r} f_{r1}\over d_{1r1}} ;\; d_{1r1} = 1 - f_{1r} f_{r1} = d_{r1r}\,.
\ee
Let us note that (3.26) is the equation for "correlation function"
$\tilde{\kappa}_{ij}^{(1)}$ of the Ornstein-Zernike type. If we take
 into account the relation $\F_{ij}^{(2)}=F_i^{(2)}\kappa_{ij}^{(1)}$
 we obtain the equation of the present type for pair CF $\F_{ij}^{(2)}$.
In the works~[32,33]
the equation of this type was constructed artificially. This fact
leads to unselfconsistent results for static susceptibility obtained
from CFs and from free energy in the case of ordered phase.
     For uniform field the following relations take place
\beea
&&f_{1r} = f \pi _{1r}; d_{1r1} = d = 1-f^2; \,
\Theta _{11} = \Theta  = {1+(z-1)f^2\over 1-f^2};  \\
&&F^{20} = {1+a^2+2a \cosh 2\beta \tilde{\tilde{\kappa}}\over [a + \cosh 2\beta \tilde{\tilde{\kappa}}]^2} ;\,
 F^{11} = {1-a^2\over [a + \cosh 2\beta \tilde{\tilde{\kappa}}]^2};  \nonumber\\
&&f = {1-a^2\over 1+a^2+a \cosh 2\beta \tilde{\tilde{\kappa}}} \stackrel{D=1}
{\rightarrow} {\cosh(\beta \kappa )-r\over \cosh(\beta \kappa )+r} \,
\eeea
Performing Fourier-transformation for (3.26) and taking into account
(3.23a) we obtain the expression for the pair CF of IM in TPCA
\be
\F^{(2)}_{(\set{q})} = F^{(2)} {1 + f\over 1-(z-1)f+(1-f)^{-1}fz\Theta (\set{q})}
= {\F^{(2)}_{(0)}\over 1+\Phi z \Theta (\set{q})}\,.
\ee
Here the following notations are introduced
\beea
&&\F^{(2)}_{(0)} = {F^{(2)}(1+f)\over 1-(z-1)f} ;\qquad \Phi  = {f(1-f)^{-1}\over 1-(z-1)f}; \\
&&\Theta (\set{q}) = {1\over z} [\pi (0) - \pi (\set{q})] = {2\over D} \sum^D_{i=1} \sin ^2(q_i/2) ;   \nonumber\\
&&\pi (\set{q}) = \sum_r \pi (\set{r}) e^{-i\set{q}\set{r}} = 2 \sum^D_{i=1} \cos  q_i \,.
\eeea
Let us note that equation (3.26) in site space is exact for the tree-type
lattice. But for Fourier-transformation (3.30) we use the translational
symmetry. Thus the expression (3.29) is exact only for 1D case. It leads to
the well known results, which we write in both wave and site spaces
($\set{q}=\set{q_1}$)~[46-48]
\be
\F^{(2)}_{(q)} = {\cosh \beta  \kappa \, a^2\, r^{-1}\over r^2+(1-a^2)\sin ^2 q/2} ;
\qquad \ang{S_0S_n} = \pmatrix{{a\over r}}^2 f^n \,.
\ee
The expression for $\ang{S_0S_n}$ is also obtained in the work~[35] in
another way.

        Now we are going to consider how to obtain the equation for
$\kappa_{ijl}^{(2)}$. It will give us the possibility to calculate the
ternary  CF. Differentiating (3.26) one finds
\bee
&&\Theta _{11} \tilde{\kappa }^{(2)}_{123} =- \sum_r {f^{(1)}_{1r3} f_{r1} + f_{1r} f^{(1)}_{r13}\over d^2_{1r1}} \tilde{\kappa }^{(1)}_{12} +  \nonumber\\
&&+ \sum_r {f^{(1)}_{1r3} + f^2_{r1} f^{(1)}_{r13}\over d^2_{1r1}} \tilde{\kappa }^{(1)}_{r2} + \sum_r {f_{1r}\over d_{1r1}} \tilde{\kappa }^{(2)}_{r23}\,.
\eee
Here the following notations are used
\be
f^{(1)}_{1r3} = {\delta \over \delta \kappa _3} f_{1r} ;\qquad f^{(1)}_{r13} = {\delta \over \delta \kappa _3} f_{r1}\,.
\ee
The derivatives (3.33) one can easily compute:
\bee
&f^{(1)}_{1r3} = f^{(01)}_{1r} d^{-1}_{1r1} [\tilde{\kappa }^{(1)}_{r3} - f_{r1} \tilde{\kappa }^{(1)}_{13}],  \nonumber\\
&f^{(1)}_{r13} = f^{(01)}_{r1} d^{-1}_{1r1} [\tilde{\kappa }^{(1)}_{13} - f_{1r} \tilde{\kappa }^{(1)}_{r3}],
\eee
where
\bee
&&f^{(01)}_{ij} = {\partial\over \partial_j\tilde{\kappa }_i} f_{ij} = - 2f_{ij} [f^{(01)}_{ij} - f_{ij} F^{(10)}_{ij}], \nonumber\\
&&f^{(10)}_{1r} = {\partial\over \partial_r\tilde{\kappa }_1} f_{1r} = 0;\qquad f^{(10)}_{r1} = {\partial\over \partial_1\tilde{\kappa }_r} f_{r1} = 0.
\eee
Substituting (3.35) into (3.32) one can find the equation for ternary "CF"
$\kappa_{ijl}^{(2)}$
\bee
&&\Theta _{11} \tilde{\kappa }^{(2)}_{123} = \sum_r {f_{1r}\over d_{1r1}} \tilde{\kappa }^{(2)}_{r23}\\
&&+\sum_r\left\{A_{1r} \tilde{\kappa }^{(1)}_{12} \tilde{\kappa }^{(1)}_{13}
+ B_{1r} [ \tilde{\kappa }^{(1)}_{12} \tilde{\kappa }^{(1)}_{r3} +
 \tilde{\kappa }^{(1)}_{13} \tilde{\kappa }^{(1)}_{r2} ] - C_{1r} \tilde{\kappa }^{(1)}_{r2} \tilde{\kappa }^{(1)}_{r3}\right\}\,.\nonumber
\eee
Here the notations are introduced
\bee
&A_{1r} = [ f^{(01)}_{1r} f^2_{r1} - f^{(01)}_{r1} f_{1r} ] d^{-3}_{1r1}\,, \nonumber\\
&B_{1r} = [ f^{(01)}_{r1} f^2_{1r} - f^{(01)}_{1r} f_{r1} ] d^{-3}_{1r1}\,, \\
&C_{1r} = [ f^{(01)}_{r1} f^3_{1r} - f^{(01)}_{1r} ] d^{-3}_{1r1}\,.\nonumber
\eee
Taking into consideration (3.28) the equation (3.36) for uniform field can
be written in the form
\bee
&&\Theta  \tilde{\kappa }^{(2)}_{123} = {f\over d} \sum_r \pi _{1r} \tilde{\kappa }^{(2)}_{r23}\\
&&+\sum_r \pi _{1r}\left\{A [ \tilde{\kappa }^{(1)}_{12} \tilde{\kappa }^{(1)}_{13}
+ \tilde{\kappa }^{(1)}_{12} \tilde{\kappa }^{(1)}_{r3} + \tilde{\kappa }^{(1)}_{13} \tilde{\kappa }^{(1)}_{r2} ] -
C_{1r} \tilde{\kappa }^{(1)}_{r2} \tilde{\kappa }^{(1)}_{r3}\right\}\,,\nonumber
\eee
where
\be
A = {2mf^2\over (1+f)^2d} ;\qquad C = {2mf(1+f+f^2)\over (1+f)^2d}\,.
\ee
Note that (3.38) is the three-particle analogy of Ornstein-Zernike equation.
Performing Fourier-transformation in (3.38)   and using the relation
\be
\Theta - {f\over d} \pi (\set{q}_1 + \set{q}_2) = {1\over \tilde{\kappa }^{(1)}(\set{q}_1+\set{q}_2)}
\ee
we get the following result for $\tilde{\kappa}^{(2)}(\set{q}_1,\set{q}_2)$
\bee
\tilde{\kappa }^{(2)}(\set{q}_1\set{q}_2) =\left\{[z + \pi (\set{q}_1) + \pi (\set{q}_2)] A-C \pi (\set{q}_1+\set{q}_2)\right\}\times \nonumber\\
\times \tilde{\kappa }^{(1)}(\set{q}_1) \tilde{\kappa }^{(1)}(\set{q}_2) \tilde{\kappa }^{(1)}(\set{q}_1+\set{q}_2).
\eee
Substituting (3.41) in (3.23b) for ternary CF
$\!\!F^{(3)}(\set{q}_1,\set{q}_2)$ we finally find~[35]

\bee
\F^{(2)}(\set{q}_1\set{q}_2) =\left\{F^{(3)} + F^{(2)} [(z+\pi (\set{q}_1) + \pi (\set{q}_2)) A -\right. \nonumber\\
\left. - C \pi (\set{q}_1+\set{q}_2)] \tilde{\kappa }^{(1)}(\set{q}_1+\set{q}_2)\right\}\tilde{\kappa}^{(1)}(\set{q}_1) \tilde{\kappa }^{(1)}(\set{q}_2).
\eee
It is easy to see (invoking (3.39)) that at $T>T_c$ the relation
$\F^{(3)}(\set{q}_1,\set{q}_2)=0$ fulfils. For computation of
$\kappa_{123}^{(2)}$, $\F_{123}^{(3)}$ it is necessary to transform (3.41),
(3.42) to site representation. The expressions for them (see~[35]) will
contain the sums over sites. In the 1D case these sums due to the form
$\kappa_{12}^{(1)}=f^{|\set{1}-\set{2}|}$ lead to the sums of
geometrical progression. After cumbersome transformation the following
expression for ternary CF can be obtained~[35]
\be
F^{(3)}(j,k) = \ang{S_0S_jS_k}^c = F^{(3)} f^k;\;
\tilde{\kappa }^{(2)}(j,k) = 2m f^k(f^j - 1);\;0\le j\le k.
\ee
It is important to note that $\F^{(3)}(j,k)$ does not depend on intermediate
index $j$ ($0\le j\le k$). Let us note that the expression for $\F^{(3)}(j,k)$
which is obtained by us in the frame of TPCA coincides with the result of
works~[47,48]. In these works the explicit expressions for mentioned above quantities
are written only at $k=j+1$, $j+2$. In the work~[35] using the solution for
the fields $_r\varphi^{(2)}(j,k)$ the expression (3.43) is obtained
immediately in the site representation.

\subsection{Relaxational dynamics.}

        In the present subsection we shall consider the relaxational dynamics
of the IM. Our main task is the calculation of the longitudinal dynamic
susceptibility of the system under consideration
\bee
&&\chi (\set{q},\omega ) =\int_{-\infty}^{+\infty} dt e^{-{\it i}\omega t} \sum_{i-j} e^{{\it i} \set{q}\set{R}_{ij}} \chi _{ij}(t)\,, \nonumber\\
&&\chi_{ij}(t,t') = {\delta  m_{\myhbox{it}}\over \delta  \kappa _{\myhbox{jt'}}}|_{\delta \kappa _{\myhbox{jt'}}=0};\; m_{\myhbox{it}} = \ang{S_i}_{\rho (t)} = m_i + \delta  m_{\myhbox{it}}\,.
\eee
Here we consider the dynamical response of the system with respect to small
time-dependent mechanical perturbation of the initial Hamiltonian in the form
$\sum_j \delta \kappa_i(t) S_j^z$. For the system described by the
Hamiltonian (3.1) we can calculate only the static susceptibility
$\chi(\set{q},0)$. The evaluation of dynamic susceptibility can be
accomplished if we shall take aditional to (3.1) items which contain other
component of pseudospins or if we shall take into account the interaction of the IM
with dissipative subsystem (thermostat)~[48]. Further we shall investigate the
system which contains becides the (3.1) with the substitution
$\kappa_i\rightarrow\kappa_{i,t}=\kappa_i+\delta\kappa_{i,t}$ also the
Hamiltonian of thermostat and interaction of thermostat with the Ising
subsystem in the form $\sum_{ja}U_j^aS_j^a$ ($a=x,y,z$). Here $U_j^a$
contains only the variables of thermostat. Within the framework of
nonequilibrium statistical operator (NSO) method one can obtain~[48-51] the
kinetic equation (KE) for unary CF $\ang{S_i}_{\rho(t)}$. Using some
approximations~[48] this KE can be lead to the Glauber-type form~[52,46]
\beea
&&[1 + \tau_i \Omega _{\myhbox{it}}] \ang{S_i}_{\rho (t)} = \ang{F^{(1)}(\epsilon _{\myhbox{it}})}_{\rho (t)}\,, \\
&&\Omega _{\myhbox{it}} = {d\over dt} + \zeta_i/\tau_i {d^2\over dt^2} ;\qquad \epsilon _{\myhbox{it}} = \kappa _{\myhbox{it}} + \sum_r K_{1r} S_r\, .
\eeea
Here $\tau_i=\gamma_i^{-1}$ is the bare time of relaxation ($\gamma_i$ is the
bare damping) which depends on the parameters of thermostat and on
parameters of interaction of pseudospin subsystem with the thermostat.
In addition to~[48] we introduce also the bare resonance
frequency $\omega_i$ ($\zeta_i=\omega_i^{-2}$). This is entirely in the
spirit of phenomenological equation for damped oscilator (see~[5]) with
frequency $\omega_i$ and with the damping $\gamma_i$. The equation (3.45a) in
the case $\zeta_i=0$ can be obtained also if we shall proceed from the Glauber
master equation for the density matrix of the system~[52,46]. If we shall develop
$F^{(1)}(\epsilon_{it})$ as a set in $S_r$ one can obtain expression for
$\ang{F^{(1)}(\epsilon_{it})}$ which contains distribution functions
starting from unary $\ang{S_i}_{\rho(t)}$ up to the function of the power $z$
$\ang{S_{i_1}S_{i_2}\cdot\cdot\cdot S_{i_z}}_{\rho(t)}$. The site indices of
mentioned above functions belong to the certain coordinate sphere $\pi_i$
of the site $i$. In the following for average
$\ang{F^{(1)}(\epsilon_{it})}$ we shall restrict ourselves only by one-particle
approximation
\beea
&&\ang{F^{(1)}(\epsilon _{\myhbox{it}})}_{\rho _{(t)}}\approx F^{(1)}(\tilde{\kappa }_{\myhbox{it}})\,, \\
&&\tilde{\kappa }_{\myhbox{it}} = \kappa _{\myhbox{it}} + \Delta _{\myhbox{it}} ;\; \Delta _{\myhbox{it}} = \sum_{r\in \pi _i} \,_r\varphi _{\myhbox{it}} = \sum_{r\in \pi _i} K_{ir}\;^{\varphi} \ang{S_r}_t
\eeea
and two-particle one
\beea
&&\ang{F^{(1)}(\epsilon _{\myhbox{it}})}_{\rho _{(t)}}\approx\ang{F^{(1)}(_j\tilde{\kappa }_{\myhbox{it}} + K_{ij} S_j)}_{\rho _{(t)}} \equiv \nonumber \\
&&\equiv  L_i(_j\tilde{\kappa }_{\myhbox{it}} , K_{ij}) + P(_j\tilde{\kappa }_{\myhbox{it}} , K_{ij}) \;^k\ang{S_j}_t \,, \\
&&_j\tilde{\kappa }_{\myhbox{it}} = \tilde{\kappa }_{\myhbox{it}} - \,_j\varphi _{\myhbox{it}} = \kappa _{\myhbox{it}} + \sum_{r\in \pi_i;r\neq j} \,_r\varphi _{\myhbox{it}}\,.
\eeea
Here the following notations are introduced
\beea
&&L(_j\tilde{\kappa }_{\myhbox{it}}, K_{ij}) = L_{ij,t} = {1\over 2} \sum_{\sigma =\pm 1} F^{(1)}(_j\tilde{\kappa }_{\myhbox{it}} + K_{ij}\sigma ) \,,\\
&&P(_j\tilde{\kappa }_{\myhbox{it}}, K_{ij}) = P_{ij,t} = {1\over 2} \sum_{\sigma =\pm 1} \sigma  F^{(1)}(_j\tilde{\kappa }_{\myhbox{it}} + K_{ij}\sigma ).
\eeea
The field $_r\varphi_{it}$ is taken to be the same in both approximations.
The equations for $_r\varphi_{it}$ (or for $\ang{S_i}_t$) will be
written below. The direct evaluation gives the following results
($a_{ij}=\exp(-2\beta K_{ij})$):
\beea
&&L_{ij,t} = {\sinh[2\beta \,_j\tilde{\kappa }_{\myhbox{it}}]\over \cosh[2\beta \,_j\tilde{\kappa }_{\myhbox{it}}] + \cosh2\beta K_{ij}}=\nonumber\\
&& = {2a_{ij} \sinh[2\beta \,_j\tilde{\kappa }_{\myhbox{it}}]\over 1+a^2_{ij}+2a_{ij} \cosh[2\beta \,_j\tilde{\kappa }_{\myhbox{it}}]}\,,  \\
&&P_{ij,t} = {\sinh 2\beta  K_{12}\over \cosh[2\beta \,_j\tilde{\kappa }_{\myhbox{it}}] + \cosh2\beta K_{ij}}=\nonumber\\
&& = {1-a^2_{ij}\over 1+a^2_{ij}+2a_{ij} \cosh[2\beta \,_j\tilde{\kappa }_{\myhbox{it}}]} \,,\\
&&P_{ij,t} = {F^{(11)}(_j\tilde{\kappa }_{it},\,_i\tilde{\kappa }_{jt})\over F^{(20)}(_j\tilde{\kappa }_{it},\,_i\tilde{\kappa }_{jt})} \mid_{\kappa_{ij}=\kappa}\rightarrow f\,.
\eeea
In the last relation we take into account the expressions for intracluster
CF $F^{(11)}_{ij}$, $F^{(20)}_{ij}$ (see (3.28a)). Substituting (3.46a) and
(3.47a) at $i=1,2$ in (3.45a) and writing two-site systems
of equations in the matrix form we obtain ($^1m_{it}=\,^1\ang{S_i}_t$;
$^km_{it}=\,^k\ang{S_i}_t$)
\beea
&&\tau_i \Omega _{\myhbox{it}} \,^1m_{\myhbox{it}} = F^{(1)}(\tilde{\kappa }_{\myhbox{it}}) - \,^1m_{\myhbox{it}} ; \nonumber\\
&&\hat{d}_t \,^k{\bf m}_t = {\bf L}_t - \hat{{\cal P}}_t \,^k{\bf m}_t ;\qquad \hat{d}_t = \{\delta _{ij} \tau_i \Omega _{\myhbox{it}}\}; \\
&&^k{\bf m}_t = \pmatrix{^k{{\bf m}_{1t}}\cr ^k{{\bf m}_{2t}}};
{\bf L}_t =\pmatrix{L_{12,t}\cr L_{21,t}};
\hat{{\cal P}}_t =\pmatrix{1&-P_{12,t}\cr -P_{12,t}&1}\,.
\eeea
By the direct computation one can obtain the following relations
\bee
&&{\bf F}^{10}_t = {\cal P}^{-1}_t {\bf L}_t ;\qquad { D}_t = 1 - P _{12,t}\,P _{21,t}\,; \nonumber\\
&&\pmatrix{F^{10}_{12,t}\cr F^{10}_{21,t}} = {1\over { D}_t}\pmatrix{L_{12,t}+P_{12,t}\;L_{21,t}\cr L_{21,t}+P_{21,t}\;L_{12,t}}\,.
\eee
Here the function
$F^{(10)}_{12,t}=F^{(10)}(\,_2\tilde{\kappa}_{1t},\,_1\tilde{\kappa}_{2t})$ already
was used by us in the subsection~3.2. The system (3.50) can be represented in
the form
\beea
&&\tau_1 \Omega _{\myhbox{it}} ^1m_{\myhbox{it}} = F^{(1)}_{i,t} - ^1m_{i,t}\,, \\
&&{ D}^{-1}_t[\tau_1 \,\Omega _{\myhbox{it}} \,^km_{\myhbox{it}}
+ {\cal P}_{ij,t}\, \tau_j\, \Omega _{jt} \,^km_{jt}] = F^{10}_{ij,t} - ^km_{i,t}\,.
\eeea
The dynamic version of TPCA is constructed similarly to the static case.
Thus the equality $^1m_{it}=\,^km_{it}$ gives the equation for the fields
$_j\varphi_{it}$.

        For the static fields $\kappa_{it}=\kappa_i$ the left side of
(3.52) is equal to zero and we obtain the system of equations (3.8) for the
nonfluctuating part $m_i$. Let us represent the effective fields in
the form of sum of static and fluctuating parts.
\beea
&&\kappa _{\myhbox{it}} = \kappa _i + \delta  \kappa _{\myhbox{it}} ;\qquad _j\varphi _i = \,_j\varphi _i + \delta \,_j\varphi _{\myhbox{it}}\,; \\
&&\tilde{\kappa }_{\myhbox{it}} = \tilde{\kappa }_i + \delta  \kappa _{\myhbox{it}} ;\qquad _j\tilde{\kappa }_{\myhbox{it}} = \,_j\tilde{\kappa }_i + \delta \,_j\tilde{\kappa }_{\myhbox{it}}\,; \\
&&\delta \,_j\tilde{\kappa }_{\myhbox{it}} = \delta  \tilde{\kappa }_{\myhbox{it}} - \delta \,_j\varphi _{\myhbox{it}} ;\qquad \sum_{j\in \pi _i} \delta \,_j\varphi _{\myhbox{it}} = \delta  \tilde{\kappa }_{\myhbox{it}} - \delta  \kappa _{\myhbox{it}}\,.
\eeea
Let us substitute (3.53) in (3.52) and obtain two systems of equations. The
first system of equations for $m_i$ coincides with (3.8) and the second one
for $\delta m_{it}$ has the form
\beea
\tau_i \Omega _{\myhbox{it}} \delta  m_{\myhbox{it}} = F^{(2)}_i \delta  \tilde{\kappa }_{\myhbox{it}} - \delta  m_{\myhbox{it}} \,,\\
{ D}^{-1}_t\,[\tau_1\, \Omega _{\myhbox{it}} \,\delta  m_{\myhbox{it}} + { P}_{ij}\, \tau_j \,\Omega _{jt}\, \delta m_{jt}] = \nonumber\\
= F^{(20)}_{ij}[\delta  \tilde{\kappa }_{\myhbox{it}} - \delta \,_j\varphi _{\myhbox{it}}] + F^{(11)}_{ij} [\delta  \tilde{\kappa }_{jt} - \delta\, _i\varphi _{jt}] - \delta  m_{i,t}\,.
\eeea
Here the fact that in TCPA $\delta m_{it}=\delta ^1m_{it}=\delta ^km_{it}$
is taken into account. Using (3.54a) the terms of type
$\tau_i\omega_{it}\delta m_{it}$ can be excluded from (3.54b). As the result
we obtain the relationship which connects $\delta_j\varphi_{it}$,
$\delta_i\varphi_{jt}$, $\delta\kappa_{it}$, $\delta m_{it}$. Just one more analogous
 equation can be obtained by permutation of indices
$i\leftrightarrow j$. From this system we express the field $\delta
_j\varphi_{it}$ via $\delta\kappa_{it}$ and $\delta m_{it}$. Using relation
(3.53b) we obtain the equation which one can write together with (3.54a) in
the case of uniform field in the following form ($\tau_i=\tau_G$)
\beea
&&F^{(20)} \delta  \tilde{\kappa }_{\myhbox{it}} = [1 + \tau_G \,\Omega _t] \delta  m_{it}\,, \\
&&\left[1 + (z-1)f^2 F^{(20)}\right] \delta  \tilde{\kappa }_{\myhbox{it}} =\nonumber\\
&&=F^{(20)}(1-f^2)\delta  \kappa _{\myhbox{it}} + f \sum_{r\in \pi _i} \pi _{ir} \delta  m_{rt}\,.
\eeea
Substituting $F^{(20)}\delta\hat{\kappa}_{it}$ from (3.55a) in (3.55b) we
obtain the closed set of equations for $\delta m_{it}$. Differentiating the
last equation with respect to $\kappa_{jt}$ and using (3.44) we find the
expression for dynamic susceptibility of the reference system
\be
\chi (\set{q},\omega ) = {\chi (\set{q})\over 1 + \Omega (\omega ) \tau(\set{q})} ;
\; \Omega (\omega ) = i\omega [1 + {\zeta\over \tau_G} i\omega ],
\ee
where $\chi(\set{q})=\F^{(2)}(\set{q})$ (see (3.29), (3.30)) and
\be
\tau(\set{q}) = {\tau(0)\over 1 + \Phi  z \Theta (\set{q})} ;
\; \tau(0) = \tau_G {1 +(z-1) f^2\over (1-f) [1-(z-1)f]}\,.
\ee
In the 1D case ($z=2$, $J(q)=0$) when the fields $\kappa$ are
absent ($\kappa=0$) from (3.56) (at $\zeta=0$) one obtains the expression
which coincides with the well-known exact solution of kinetic equation
(3.45) (see~[46,53]). This one gives ($a=\exp(-2\beta K)$, $q=q_1$)
\be
\chi (\set{q}) = {a\over a^2+(1-a^2)\sin ^2q/2} ;\qquad \tau(\set{q}) = {\tau_G\over 2} {1 + a^2\over a^2+(1-a^2)\sin^2q/2}\,.
\ee
In the work~[54] the cluster solution of (3.45) ($\zeta=0$) for
$\chi(\set{q},\omega)$ at arbitrary value of $z$ was presented only at
$\kappa=0$, $\set{q}=0$, and this result coincides with (3.56).
   Let us now take into account the long-range interaction within MFA. Now,
as before, the equation for $m_{it}=\ang{S_i}_{\rho(t)}$ has the form
(3.45a) but we use the system (3.55) in which $\delta \kappa_{it}=\delta
h_{it}+\sum_kJ_{ik}\delta m_{kt}$. We find the susceptibility $^k\chi(q,\omega)$
with the aid of derivative $^L\chi_{ij}(t-t')=\delta m_{it}/\delta
h_{it'}$ at $\delta h_{it}=0$. This one leads to the relations of (2.7) and
(3.56) type
\be
^L\chi (\set{q},\omega ) = {\chi (\set{q},\omega )\over 1 - J(\set{q}) \chi (\set{q},\omega )} = {^L\chi (\set{q})\over 1 + \Omega (\omega )\, ^L\tau(\set{q})}\,.
\ee
In the expressions for $^L\chi(\set{q})$ and $^L\tau(\set{q})$ it is
convenient to separate the part that is independent on wave vector
\bee
&&^L\chi (\set{q}) = {^L\chi (0)\over 1 + \,^L\Phi (\set{q})} ;\qquad ^L\tau(\set{q}) = {^L\tau(0)\over 1 + \,^L\Phi (\set{q})}\,; \nonumber\\
&&^L\Phi (\set{q}) = \,^L\Phi  z \Theta (\set{q}) + \,^L\chi (0) [J_0 - J(\set{q})] ;\; J_0 = J(\set{q} \rightarrow  0)\,,
\eee
where
\bee
&&^L\chi (0) = {F^{20}(1+f)\over {\cal D}} ;\; ^L\tau(0 = {1+(z-1)f^2\over (1-f) {\cal D}} ;\; ^L\Phi  = {f\over (1-f){\cal D}}; \nonumber\\
&&{\cal D} = 1 - (z-1) f - J_0 F^{20}(1+f)\,.
\eee
        If $\set{q}\rightarrow 0$ the expressions (3.60) coincide with the
results of the works~[55,56].

\setcounter{equation}{0}
\section{Application of the two-particle cluster approximation to the Ising
model with the arbitrary value of spin.}

In the present section we shall consider the Ising-type model with the
arbitrary value of spin $M$ taking into account only pair interaction
($W(\{S\})=0$). The series of concrete results we shall obtain at $M=2$ for
the Blume-Emery-Griffits (BEG) model:
\be
H(\{\kappa \} = \sum_i [\kappa_i S_i + \kappa_i' S^2_i] + \sum_{(ij)} [K S_iS_j + K' S^2_i S^2_j]
\ee
which was suggested in the work~[57] for the investigations of phase
transition in mixture of $He^3$-$He^4$. This model corresponds to the
system with the Hamiltonian (2.1) at $M=2$ in the case
$K_{ij}^{(\mu\nu)}=\delta_{\mu\nu}K^{(\mu)}\pi_{ij}$,
$J_{ij}^{(\mu\nu)}=\delta_{\mu\nu}J^{(\mu)}$. In what follows below we do
not write the index 1 and
instead of index 2 we use the prime symbol. Let us note, that at
$\kappa'=-\frac{8}{3}zK'$ the Hamiltonian (4.1) describes the
system with the bilinear interaction $KS_iS_j$, with the quadrupolar
interaction $K'(S_i^2-8/3)(S_j^2-8/3)$ and with the external field
$\kappa$~[2]. Let us note that thermodynamic properties of the BEG model
were investigated by various methods~[2] including TPCA. However the
calculation of correlattors have not been carried out untill present time.
We will computate here within TPCA the pair CFs for the IM with arbitrary
value of spin~[37].

As it was noted in TCPA we can use the relation (2.22)-(2.26)
with $s_i=z$, $b_{ff'}=1$, $W(\{R\})=0$, $R=(12)$. Now the ${\cal F}$-function of
the system under consideration takes the form
\be
\F(\{\kappa ^{(\cdot )}\}) = (1-z) \sum^{}_1 F_1(\{\tilde{\kappa }^{(\cdot )}_1\}) + {1\over 2} \sum^{}_{1,r} F_{1r}(\{_r\tilde{\kappa}^{(\cdot )}_1\},\{_1\tilde{\kappa }^{(\cdot )}_r\})\,.
\ee
The one-particle Hamiltonian and two-particle one in the case of BEG model
read~[37]:
\beea
&&{H}_1(S_1) = \tilde{\kappa }_1S_1 + \tilde{\kappa }'S^2_1\,, \nonumber\\
&&\tilde{\kappa }_1 = \kappa _1 + \sum_{r'\in \pi _1} \,_{r'}\varphi _1; \; \tilde{\kappa }_1'= \kappa_1'+ \sum^{}_{r'\in \pi _1} \,_{r'}\varphi_1'\,, \\
&&{H}_{12}(S_1S_2) = _2\tilde{\kappa }_1S_1 + _2\tilde{\kappa }_1S^2_1 + _1\tilde{\kappa }_2S_2 + _1\tilde{\kappa }_2S^2_2 + K S_1S_1 +\myhbox{ K'S}^2_1S^2_2\,, \nonumber\\
&&_2\tilde{\kappa }_1 = \tilde{\kappa }_1 - _2\varphi _1 ;\qquad _2\tilde{\kappa}_1= \tilde{\kappa }_1- _2\varphi_1'\,.
\eeea
Let us write the expression for intracluster ${\cal F}$-function of BEG model
\beea
&&F_1(\tilde{\kappa }_1, \tilde{\kappa }_1') = \ln  2[e^{4\tilde{\kappa}_1'} \cosh 2 \tilde{\kappa }_1 + 1/2]\,,  \\
&&F_{12}(\,_2\tilde{\kappa }_1, \,_2\tilde{\kappa }_1'\mid \,_1\tilde{\kappa }_2, \,_1\tilde{\kappa }_2') =
\ln  2\left\{ e^{4(\,_2\tilde{\kappa }_1'+\,_1\tilde{\kappa }_2')+16K'}\right. \nonumber\\
&&[e^{4K} \cosh 2 (\,_2\tilde{\kappa }_1 + \,_1\tilde{\kappa }_2)
+ e^{-4K} \cosh 2 (\,_2\tilde{\kappa }_1 - \,_1\tilde{\kappa }_2)] +\nonumber\\
&&\left.+ e^{4\,_2\tilde{\kappa }_1'} \cosh 2 \,_2\tilde{\kappa }_1
 + e^{4\,_1\tilde{\kappa }_2'} \cosh 2 \,_1\kappa _2 + 1/2\right\}\,.
\eeea
Similar to (3.5) we shall use the notations for derivatives
\beea
&&\tilde{\kappa }\pmatrix{\mu_1&\mu_2\cr i_1&i_2}= \delta _{i_1i_2} \delta _{\mu _1\mu _2} + \sum^{}_{r\in \pi _{i_1}} \,_r\varphi \pmatrix{\mu_1&\mu_2\cr i_1&i_2}\,, \nonumber\\
&&\tilde{\kappa }\pmatrix{\mu_1&\mu_2\cr i_1&i_2}= {\delta \over \delta  \kappa ^{(\mu _2)}_{i_2}} \tilde{\kappa }^{(\mu _1)}_{i_1} ;\;
_r\varphi \pmatrix{\mu_1&\mu_2\cr i_1&i_2}= {\delta \,_r\varphi ^{(\mu _1)}_{i_1}\over \delta  \kappa ^{(\mu _2)}_{i_2}}\,, \\
&&F^{(k_\nu\nu+k_\mu\mu)}_1 (\{\tilde{\kappa }^{(\cdot )}_1\}) = {\partial^{k_\mu}\over \partial [\tilde{\kappa }^{(\mu)}_1]^{k_\mu}}
 {\partial^{k_\nu}\over \partial [\tilde{\kappa }^{(\nu)}_1]^{k_\nu}} F_1(\{\tilde{\kappa }^{(\cdot )}\}) \,,\nonumber\\
&&F^{(k_\mu\mu+k_\nu\nu\mid l_\delta\delta+
l_\epsilon\epsilon)}_{1r} (\{\,_r\tilde{\kappa }^{(\cdot )}_1\}\mid \{\,_1\tilde{\kappa }^{(\cdot )}_r\}) = \\
&&{\partial^{k_\mu}\over \partial [\,_r\tilde{\kappa }^{(\mu)}_1]^{k_\mu}}
{\partial^{k_\nu}\over \partial [\,_r\tilde{\kappa }^{(\nu)}_1]^{k_\nu}}
{\partial^{l_\delta}\over \partial [\,_1\tilde{\kappa }^{(\delta)}_r]^{l_\delta}}
{\partial^{l_\epsilon}\over \partial [\,_1\tilde{\kappa }^{(\epsilon)}_r]^{l_\epsilon}}
F_{1r}(\{\,_r\tilde{\kappa }^{(\cdot )}_1\}\mid \{\,_1\tilde{\kappa }^{(\cdot )}_r\})\,. \nonumber
\eeea
Using the relations (2.5) and stationarity condition
$\dif{}{\varphi_i^{(\mu)}}F(\{\kappa^{(\cdot)}\})=0$ one can find the system
of $N(z+1)M$ equations in $NZ$ unknowns $\ang{S_1^\mu}$ and $NzM$ unknowns
$_r\varphi^{\mu}_i(\{\kappa^{(\cdot)}\}, \{K^{(\cdot)}\})$
\beea
&&\F^{(1)}\left(^{\mu_1}_1;\{\kappa^{(1)}\}\right) = \ang{S^{\mu_1}_1}_\rho = \ang{S^{\mu_1}_1}_{\rho_1} = F^{(\mu _1)}_1(\{\tilde{\kappa}^{(\cdot )}_1\}) \,,\\
&&F^{(\mu_1)}_1(\{\kappa^{(\cdot )}_1\}) = \ang{S^{\mu_1}_1}_{\rho_1} =\nonumber\\
&&= \ang{S^{\mu_1}_1}_{\rho_{1r}} = F^{(\mu _1|0)}_{1r}(\{_r\kappa ^{(\cdot )}_1\}\mid\{_1\kappa ^{(\cdot )}_r\})\,.
\eeea
When the long-range interaction is absent (4.6a) gives the system of
equations for $_r\varphi_i^{(\mu)}$ while (4.6a) gives the expression for
unary correlation function $\ang{S_1^\mu}_\rho$.

        The system (4.6b) is identified to $Nz$ independent relations
between trial one-particle $\rho_1(S_1)$ and two-particle
$\rho_{1r}(S_1,S_r)$ density matrices (see (3.9)). These relations and some
other ones in which the quantities $\rho_1$, $\rho_{1r}$ and $U_{1r}$ have
not explicit form remain the same form as in the case $M=1$ (see
(3.11)-(3.13)). In particular, in the expansion of ${\cal F}$-function the
$K$-function for the graph of (3.12) type give zero contribution.
Similar to the case $M=1$ the TCPA gives exact results for responces of
the IM with arbitrary $M$ on the tree-lattice. Let us note that below we
shall use the relation of (3.10) type which we can write in the form
\be
\ang{(S_1)^{K_1}\ldots (S^M_1)^{K_M} }^c_{\rho _1} =
\ang{(S_1)^{K_1}\ldots (S^M)^{K_M} }^c_{\rho _{1r}}\,.
\ee
Further we shall present the explicit form for series of responses of BEG
model in TCPA over short-range interaction and in MFA over long-range
interaction~[37]. For the free energy function per site from (2.3) and
(4.2)-(4.4) we find
\beea
&&\beta\;^Lf = {\beta \over 2} \left[J_0 m^2+J_0'q^2\right] + \beta f(\kappa ,\kappa ',K,K') \,,\\
&&-\beta f = (1-z) \ln  Z_1(\tilde{\kappa },\tilde{\kappa }') + {z\over 2} \ln  Z_{12}(\tilde{\tilde{\kappa}},\tilde{\tilde{\kappa}}')\,.
\eeea
Here the notations are used
\beea
&&Z_1(\tilde{\kappa },\tilde{\kappa }') = 2 e^{4\beta \tilde{\kappa }'} \cosh 2\beta \tilde{\kappa } + 1 \,,\\
&&Z_{12} = 2e^{\beta (8\tilde{\tilde{\kappa}}'+16K'+4K)} \left[ \cosh 4\beta \tilde{\tilde{\kappa}}+ e^{-8\beta K} \right] + \nonumber\\
&&+4e^{4\beta \kappa '} \cosh 2\beta \tilde{\tilde{\kappa}}+ 1 \,,\\
&&\tilde{\kappa } = \kappa  + z\varphi ,\;\tilde{\tilde{\kappa}}= \kappa  + (z-1)\varphi ;\qquad \kappa  = h + J_0m\,, \nonumber\\
&&\tilde{\kappa }'= \kappa ' + z\varphi',\;\tilde{\tilde{\kappa}}'= \kappa ' + (z-1)\varphi '; \qquad \kappa'  = h' + J_0'q\,, \\
&&m = \ang{S_i}, q = \ang{S^2_i}, J_0 = J(\set{q}\rightarrow 0), J_0'= J'(\set{q}\rightarrow 0)\,. \nonumber
\eeea
The system of four equations (4.6) for $m$, $q$, $\varphi$, $\varphi'$ on
the basis (4.4)-(4.5)can be obtained in the form
\beea
&&m = {4e^{4\beta \tilde{\kappa }'} \sinh 2\beta \tilde{\kappa }\over Z_1(\tilde{\kappa },\tilde{\kappa }')}\,,\\
&&q = {8e^{4\beta \tilde{\kappa }'} \cosh 2\beta \tilde{\kappa }\over Z_1(\tilde{\kappa },\tilde{\kappa }')}\,, \\
&&{e^{4\beta \tilde{\kappa }'} \sinh 2\beta \tilde{\kappa }\over Z_1(\tilde{\kappa },\tilde{\kappa }')} =\nonumber\\
&&={e^{\beta (8\tilde{\tilde{\kappa}}'+16K'+4K)} \sinh 4\beta \tilde{\tilde{\kappa}}+ e^{4\beta \tilde{\tilde{\kappa}}'} \sinh 2\beta \tilde{\tilde{\kappa}}\over Z_{12}(\tilde{\tilde{\kappa}},\tilde{\tilde{\kappa}}')}\,, \\
&&{e^{4\beta \tilde{\kappa }'} \cosh 2\beta \tilde{\kappa }\over Z_1(\tilde{\kappa },\tilde{\kappa }')} =\nonumber\\
&&={e^{\beta (8\tilde{\tilde{\kappa}}'+16K'+4K)} (\cosh 4\beta \tilde{\tilde{\kappa}}+ e^{-8\beta K})+ e^{4\beta \tilde{\tilde{\kappa}}'} \cosh 2\beta \tilde{\tilde{\kappa}}\over Z_{12}(\tilde{\tilde{\kappa}},\tilde{\tilde{\kappa}}')}\,.
\eeea
Let us note that numerical investigation of system (4.10) at different
values  of parameters of the Hamiltonian (4.1) in the case $J_0=J_0'=0$ is
carried out in the work~[37]. Here it is also investigated the
thermodynamic potential and static susceptibility of BEG model.

        The method of finding CF of IM with arbitrary value $M$ is similar
to one described in subsection~3.2. But now the matrix relation over
indices $\mu$, $\nu$ takes place. One can obtain
from the expression (4.6a) the matrix relations of (3.23) type
connecting CF with derivatives from $\tilde{\kappa}^{(\mu)}_i$.  In the present
work we restrict ourselves to
finding only pair CF. Thus differentiating  (4.6b) with respect to
$\kappa_2^{(\mu_2)}$ we obtain the relation of (3.25) type
\beea
&&_r\hat{\varphi }_{12} + \hat{f}_{1r} \,_1\hat{\varphi }_{r2} = \hat{f}_{1r}\hat{\tilde{\kappa}}_{r2}\,, \\
&&\hat{f}_{r1} \,_r\hat{\varphi }_{12} + \,_1\hat{\varphi }_{r2} = \hat{f}_{r1}\hat{\tilde{\kappa}}_{12}\,.
\eeea
Here the following notations and relations are used
\beea
&&\hat{f}_{ij} = \left[\hat{F}^{(2)}_i\right]^{-1} \hat{F}^{(11)}_{ij};\qquad \hat{F}^{(2)}_i \equiv  \hat{F}^{(20)}_{ij}\,, \\
&&\hat{F}^{(2)}_i = \left\{F^{(\mu+\nu)}(\{\tilde{\kappa }^{(\cdot )}_1\})\right\};
\hat{F}^{(11)}_{ij} = \left\{F^{(\mu \mid \nu )}(\{_j\tilde{\kappa }^{(\cdot )}_i\}\mid \{_i\kappa ^{(\cdot )}_j\})\right\}\,, \nonumber\\
&&\hat{\tilde{\kappa_{ij}}} = \left\{\kappa \pmatrix{\mu &\nu \cr i&j}\right\};
\qquad \,_r\hat{\varphi }_{ij} = \left\{_r\varphi \pmatrix{\mu &\nu \cr i&j}\right\}\,.
\eeea
So expressing the matrices $_r\hat{\varphi}_{ij}$ with the aid of
$\hat{\tilde{\kappa}}_{ij}$ and forming the closed equation for
$\hat{\tilde{\kappa}}_{ij}$ one can find at the case of uniform field the
expression for $\hat{\tilde{\kappa}}(\set{q})$. On the basis of matrix
relation of (3.23a) type we find the expression for pair CF
$\F^{(2)}(\set{q})$. It can be written in two forms~[37]
\beea
&&\hat{\F}^{(2)}(\set{q}) = \hat{F}^{(2)}\left[1-(z-1)\hat{f}+\right.\nonumber\\
&&\left.(1-\hat{f})^{-1}\hat{f}z\Theta (\set{q})\right]^{-1} (1+\hat{f})\,, \\
&& \left [\hat{\cal F}^{(2)}\right ]^{-1}(\set{q}) = (1-z)\left [\hat{F}^{(2)}\right]^{-1} + z \left[\hat{F}^{(2)}+\hat{F}^{(11)}\right]^{-1} + \nonumber\\
&&+ \left[\hat{F}^{(2)}(\hat{F}^{(11)})^{-1}\hat{F}^{(2)}-\hat{F}^{(11)}\right]^{-1} z\Theta (\set{q})\,.
\eeea
In the case of BEG model we shall present the explicit form of matrix
elements (intracluster pair CF), which are included in (4.13)
($F^{(1+1')}$, $F^{(1+0'|0+1')}$)-nondiagonal elements and $l$
($l'$)denotes the number of derivatives with respect to unprimed (primed)
field~[37]:
\beea
&&F^{(2+0')} = 8e^{4\beta \tilde{\kappa }'} \left[2e^{4\beta \tilde{\kappa }'} + \cosh 2\beta \tilde{\kappa }\right] Z^{-2}_1\,, \nonumber\\
&&F^{(0+2')} = {32e^{4\beta \tilde{\kappa }'} \cosh 2\beta \tilde{\kappa }\over Z^2_1}\,,\\
&&F^{(1+1')} = F^{(1'+1)} = {32e^{4\beta \tilde{\kappa }'} \sinh 2\beta \tilde{\kappa }\over Z^2_1} ; \nonumber\\
&&F^{(1+0'\mid 1+0')} = 8e^{8\beta \tilde{\tilde{\kappa}}'}\left\{e^{16\beta K'} \left[4e^{\beta (8\tilde{\tilde{\kappa}}'+16K')} \sinh 8\beta K +\right.\right. \nonumber\\
&&+ 8e^{4\beta \tilde{\tilde{\kappa}}'} \sinh 4\beta K \cosh 8\beta \tilde{\tilde{\kappa}}+\nonumber\\
&&\left.\left.+ e^{4\beta K} \cosh 4\beta \tilde{\tilde{\kappa}}- e^{-4\beta K}\right] - 2 \sinh^22\beta \tilde{\tilde{\kappa}}\right\}Z^{-2}_{12}\,, \\
&&F^{(0+1'\mid 0+1')} = {32\over Z^2_{12}} e^{8\beta \tilde{\tilde{\kappa}}'}\left\{e^{\beta (16K+4K)} [\cosh 4\beta \tilde{\tilde{\kappa}}+\!\! e^{-8\beta K}]\!\! -\!\! 2 \cosh^22\beta \tilde{\tilde{\kappa}}\right\}, \nonumber\\
&&F^{(1+0'\mid 0+1')}=F^{(0+1'\mid 1+0')} = {16\over Z^2_{12}} e^{8\beta \tilde{\tilde{\kappa}}'}\left\{4e^{\beta (4\tilde{\tilde{\kappa}}'+16K')} \sinh 2\beta \tilde{\tilde{\kappa}}\times\right. \nonumber\\
&&\times\sinh 4\beta K\left. + e^{\beta (16K'+4K)} \sinh 4\beta \tilde{\tilde{\kappa}}- 2\sinh 2\beta \tilde{\tilde{\kappa}}\cosh 2\beta\tilde{\tilde{\kappa}}\right\}\,.\nonumber
\eeea
Let us note that the form (4.13) is the typical one for TPCA. Formula (4.13)
gives expression for averaging over configuration of pair spin-spin CF~[58]
in the case of Ising mixture with the quenched disorder of replacement.
Now the indices $\mu$, $\nu$ number sorts of atoms; and intracluster CF
contain averaging over sort configuration. In the case of the one-sort
quantum system (the Hamiltonian of such system contains different
components of spin operators $S^a$) the form (4.13) gives expressions for the
pair cumulant temperature Green functions in the impulse-frequency
representation~[42,43]. Now the indices $\mu$, $\nu$ number the x, y, z
component of spin operators;  $\hat{F}^{(2)}\rightarrow
\hat{F}^{(2)}(\omega_n)$, $\hat{F}^{(11)}\rightarrow
\hat{F}^{(11)}(\omega_n)$ are expressed with the aid of intracluster Green
functions.

\setcounter{equation}{0}
\section{Thermodynamics and pair correlation func\-ti\-ons of
$KD_2PO_4$ ferroelectrics in four-particle cluster approximation.}

\subsection{Statement of cluster expansion method. Intracluster
cor\-re\-la\-ti\-ons function.}

	We shall consider  a system of deuterons moving on O-D..O bonds in
compound of $KD_2PO_4$ type. This compound is related to orthorombic
symmetry with the four molecules per West's elementary cell which is
characterized by basic vector: $|\set{e}_1|=|\set{e}_1|=7.469 A$,
$|\set{e}_3|=6.975A$. The unit cell of the Bravais lattice is composed of
two neighbouring $PO_4$ tetrahedra together with four hydrogen ($\alpha=$1,
2, 3, 4) bond relating to one of them (A - type tetrahedron). Hydrogen bonds
going to another (B - type) tetrahedron belong to four nearest structural
elements surrounding it. Their coordinates are $\set{n}_1$, $\set{n}_1+\set{r}_2$, $\set{n}_1+\set{r}_3$,
$\set{n}_1+\set{r}_4$. The projection of two neighbouring tetrahedra on the
plane ($\set{e}_1\times\set{e}_2$) as well as the effect of cluster fields on deuterons are
presented in the figure~2. %\\[1.5ex]

	The Hamiltonian of deuteron subsystem taking into account short-range
interaction and taking into consideration long-range interaction
$J_{\alpha\alpha}(n-n')$ in MFA can be written as follows~[32,33]
\beea
&&H(\{\kappa \}) = \sum_{n,\alpha} \kappa _{n\alpha} S_{n\alpha } - \sum_R
V_R(\{s\})\,, \\ &&V_R(\{s\}) = V(n_1, n_2, n_3, n_4) =\{ {1\over 8}
\sum_{\alpha ,\alpha '} V_{\alpha\alpha'} S_{n_\alpha\alpha}
S_{n_{\alpha'}\alpha'} + \nonumber\\ &&+ {1\over 16} \Phi  S_{n_11} S_{n_21}
S_{n_33} S_{n_44}\}\{\delta _{n_1n_2} \delta _{n_1n_3} \delta _{n_1n_4} +
\\ &&+ \delta _{n_1+r_2,n_2} \delta _{n_1+r_3,n_3} \delta
_{n_1+r_4,n_4}\}\,, \nonumber\\ &&\kappa _{n\alpha } = h_{n\alpha } + \lambda
_{n\alpha } ; \lambda _{n\alpha } = \sum_{n'\alpha '} J_{\alpha \alpha
'}(n-n') \ang{S_{n'\alpha '}}\,.
\eeea
Here $S_{n\alpha}=\pm 1$ describes the position of deuteron in double well
potential which situates on $\alpha$th hydrogen bond in $n$th
cell of Bravais lattice, $n_\alpha=n+r_\alpha$ where $r_\alpha$ is
the radius-vector of the hydrogen bond $\alpha$ in the present cell $\vec{n}$. In
(5.1b) the first term (the first product of Kroneker symbols) describes the
short-range configurational interactions of deuterons near "A" tetrahedra,
the second term describes the same interactions near "B" tetrahedra. Let us
note that parameter $J_{\alpha\alpha}(n-n')$ includes also an indirect
deuteron interaction via lattice vibration. Further we shall use the
following well-known notations
\begin{figure}
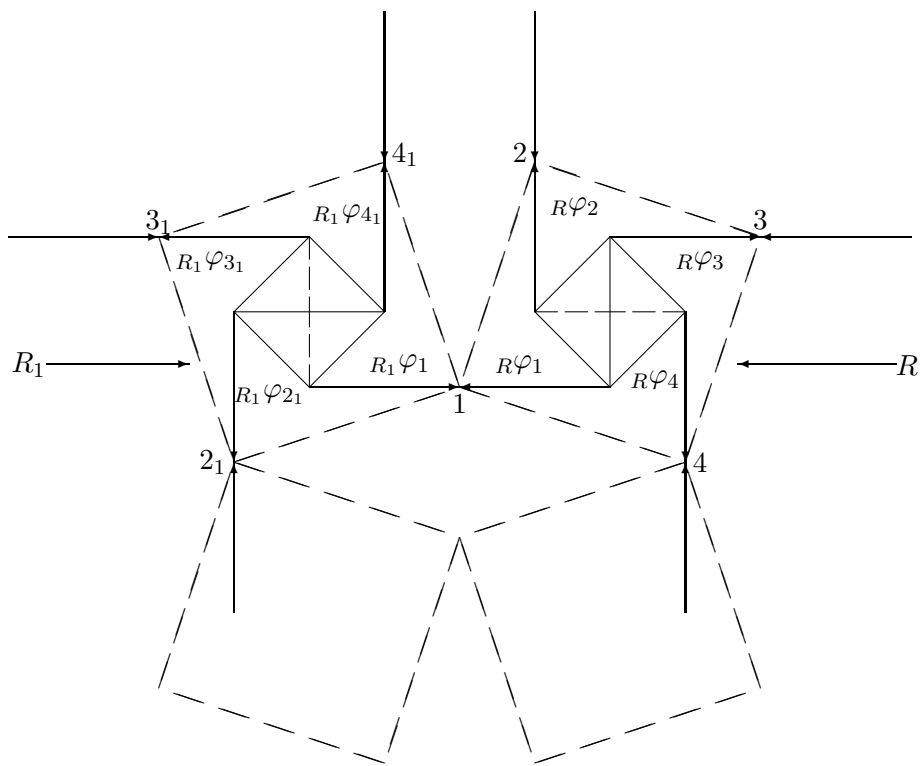

\begin{center}	\input cmpl.pic	\end{center}
\caption[]{
Two neighbouring tetrahedra (A and B type) and lattice constructed
from four-particle clusters in projection on the plane
($\set{e}_1\times\set{e}_2$). The tops of clusters coincide with the centres of
hydrogen bonds. The effect of cluster fields on sites of cluster lattice is
denoted by pointers. The sites of cluster $R_0=R(R_1)$ are numbered by indices
$i_0=i(i_1)$.
}
\end{figure}
\bee
&&V_{12} = V_{23} = V_{34} = V_{41} = - {1\over 2} W_1 = V_s \,,\nonumber\\
&&V_{13} = V_{24} = - \epsilon  + {1\over 2} W_1 = U_s\,, \nonumber\\
&&\Phi  = 4\epsilon  - 8W + 2W_1;\qquad \epsilon  = \epsilon _a - \epsilon _s \,,\\
&&W = \epsilon _1 - \epsilon _s,\qquad W_1 = \epsilon _0 - \epsilon _s\,.\nonumber
\eee
Here $\epsilon_s$, $\epsilon_1$, $\epsilon_0$, $\epsilon_a$ are four
different configurational energies for sixteen configurations of deuterons near
$[PO_4]^{3-}$ tetrahedra (see~[4,5,32]). It is convenient to use the general
index $i=(n_i,\alpha_i)$. Let us decompose pseudospin lattice into clusters.
The cluster we shall take in the form of tetrahedron. Its tops coincide with
the centres of hydrogen bonds. The projections of the same clusters on
the plane $a$x$b$ is noted by shade line (see figure~2). Let us note that the
lattice which is formed from the same clusters is similar to deformated
square lattice with chess decomposition of this lattice into square
clusters.  Evidently that the structure of the Hamiltonian (5.1) agrees with
that of (2.1). So we shall use the relations of section~2 (starting from
(2.8)) at $R=(1,2,3,4)$ when we shall perform the cluster expansion. In the
present work we limit ourselves to four-particle cluster approximation
(FPCA). Thus one may use relations (2.22)-(2.26). For reference
$\F$-function we write
\be
\F(\{\kappa \}) = - \sum_i F_i(\tilde{\kappa }_i) + \sum_R F_R(\{_R\tilde{\kappa }\})\,,
\ee
where
\beea
&&F_i(\tilde{\kappa }_i) = \ln  Z_i(\tilde{\kappa }_i) = \ln  Sp e^{\tilde{\kappa }_iS_i} = \ln  2 \cosh \tilde{\kappa }_i\,, \nonumber\\
&&F_R(\{_R\tilde{\kappa }\}) = \ln  Sp e^{ H_R(\{s\})} = F^{(0000)}_{1234} = \nonumber\\
&&=\ln  Z^{(0000)}_{1234} = F_R(_{R_1}\tilde{\kappa }_1,\; _{R_2}\tilde{\kappa }_2,\; _{R_3}\tilde{\kappa }_3,\; _{R_4}\tilde{\kappa }_4)=\nonumber\\
&& = F_R(\tilde{\tilde{\kappa}}_1,\tilde{\tilde{\kappa}}_2,\tilde{\tilde{\kappa}}_3,\tilde{\tilde{\kappa}}_4) \,,\\
&& H(\{s\}) = - V_R(\{s\}) + \sum_{f\in R} \,_{R_f}\kappa_f  S_f \,,\nonumber\\
&&\tilde{\kappa }_f = \kappa _f + \sum_R \,_R\varphi _f ;\qquad _{R_f}\kappa _f = \tilde{\kappa }_f - \,_{R_f}\varphi _f = \tilde{\tilde{\kappa}}_f \,.
\eeea
The explicit form of cluster partition function is the following
\bee
&&{1\over 2} Z_R={1\over 2} Z^{(0)}_{1234} = 2a \cosh (\tilde{\tilde{\kappa}}_1 - \tilde{\tilde{\kappa}}_3) \cosh (\tilde{\tilde{\kappa}}_2 - \tilde{\tilde{\kappa}}_4) + \nonumber\\
&&+ d \cosh (\tilde{\tilde{\kappa}}_1 -\tilde{\tilde{\kappa}}_2 + \tilde{\tilde{\kappa}}_3 - \tilde{\tilde{\kappa}}_4) + \cosh (\tilde{\tilde{\kappa}}_1 + \tilde{\tilde{\kappa}}_2 + \tilde{\tilde{\kappa}}_3 + \tilde{\tilde{\kappa}}_4) + \nonumber\\
&&+ 2b [\cosh (\tilde{\tilde{\kappa}}_1 + \tilde{\tilde{\kappa}}_3) \cosh (\tilde{\tilde{\kappa}}_2 - \tilde{\tilde{\kappa}}_4) + \\
&&+ \cosh (\tilde{\tilde{\kappa}}_1 - \tilde{\tilde{\kappa}}_3) \cosh (\tilde{\tilde{\kappa}}_2 + \tilde{\tilde{\kappa}}_4)] \,,\nonumber\\
&&a = e^{-\beta \epsilon } ;\qquad b = e^{-\beta W} ;\qquad d = e^{-\beta W_1}\,. \nonumber
\eee
Let us note that  we neglect general factor
$\exp\{(W_1+4W+2\epsilon)/8\}$ in (5.5). It gives independent of temperature
term in to free energy $-\beta^{-1}\F(\{\kappa\})$. Let us introduce the following
notations for derivatives
\bee
&&F^{(l)}_i(\tilde{\kappa}_i) = {\partial^l\over (\partial\tilde{\kappa }_i)^l} F(\tilde{\kappa }_i);
\qquad F^{(l)}_{f,R} = {\partial^l\over (\partial\tilde{\tilde{\kappa}}_f)^l} F^{(0000)}_{1234} \,,\\
&&F^{(ll')}_{\myhbox{ff'},R} = {\partial\over \partial\left(\tilde{\tilde{\kappa}}_f\right)^l} {\partial\over \partial\left(\tilde{\tilde{\kappa}}_{f'}\right)^{l'}} F^{(0000)}_{1234} ;\; f,f' = (1,2,3,4) =(R)\,.\nonumber
\eee
The system of equations for averages $m_i=\ang{S_i}$ and fields $_R\varphi_i$
is obtained as in the previous section on the basis (2.5) and stationarity
condition of $\F(\{\kappa\})$ with respect to $_R\varphi_i$ ($^Lm_i=m_i$)
\beea
&&m_i = \,^k\F^{(1)}_1 = F^{(1)}(\tilde{\kappa }_1) = \tanh \tilde{\kappa }_i \,,\\
&&F^{(1)}_1(\tilde{\kappa }_1) = F^{(1)}_{1,R_r} ;\qquad r = 0,1\,.
\eeea
Here index $\set{r}$ numbers two clusters $r=0,1$ which contain bond 1 (see
figure~2). Therefore there are two fields per one site. Then the system (5.7b)
gives $8N$ equations in $8N$ unknowns $_R\varphi_i$ where $N$ is the
number of A (or B) tetrahedron in the crystal. We have
one equation (5.7b) for the field $\varphi$ in the case of uniform field. The system (5.7b) is equivalent
to the following system of equations
\be
\rho _1(S_1) = Sp_{S_{2r}S_{3r}S_{4r}} \rho _R(S_1, S_{2r}, S_{3r}, S_{4r})\,.
\ee
In particular from (5.8) the relation between intracluster CF follows:
\be
\ang{S^l_1}^c_{\rho _1} = F^{(l)}_1(\tilde{\kappa }_1) = F^{(l)}_{1,R} = \ang{S^l_1}^c_{\rho _R}\,.
\ee
One can obtain the relation for moment function from (5.8)
\be
M({R}) = Sp_{1,2,3,4} \rho _1 \rho _2 \rho _3 \rho _4 e^{U_R} =
Sp_{2,3,4} \rho _2 \rho _3 \rho _4 e^{U_R}
\ee
One can carry out the investigation of diagrams for $F$-function and
$\ang{S_1}=\fdif{}{\kappa_1}F$ in the higher order of cluster expansion
similar as it was done in the case of two-particle cluster. This
investigation leads to conclusion that FPCA gives the exact results for all
characteristics of the system with the Hamiltonian (5.1) in the case of
tree-like lattice. This lattice is constructed on the basis of tetrahedra.
It can be formed from two-particle tree with $z=4$ if all sites will change
by tetrahedra. Now let us write some results in the homogeneous field. The
free energy per tetrahedron in FPCA taking into account long-range
interaction in MFA reads ($m_\alpha=m$):
\bee
&&f = - {kT\over N} \F(h) = {\nu _1(\hat{q}_0)\over 2} m^2 + 4kT \ln  2 \cosh \beta  \tilde{\kappa } - 2kT \ln  Z_R(\tilde{\tilde{\kappa}}) \,,\nonumber\\
&&{1\over 2} Z_R(\tilde{\tilde{\kappa}}) = 2a + d + 4b \cosh 2 \beta \tilde{\tilde{\kappa}}+ \cosh 4 \beta\tilde{\tilde{\kappa}}\,,
\eee
The eigenvalue $\nu_1(\hat{q}_0)$ of the matrix $J_{\alpha\alpha}(\hat{q}_0)$
reads:
\be
\nu _1(\hat{q}_0) = J_{11}(\hat{q}_0) + J_{13}(\hat{q}_0) + 2 J_{12}(\hat{q}_0)\,,
\ee
where $\hat{q}_0$ is the vector belong to the wave vector's star $\{\set{q}_0\}$
($\epsilon\rightarrow 0$)
\be
\{\set{q}_0\} = \pm\hat{q}_{0x}, \pm\hat{q}_{0y};\;\hat{q}_{0x} = (\epsilon ,0,0);\;\hat{q}_{0y} = (0,\epsilon ,0)\,.
\ee
Fields $\tilde{\kappa}$, $\tilde{\tilde{\kappa}}$ contain the eigenvalue
$\nu_1(\hat{q}_0)$ via molecular field $\nu_1(\hat{q}_0)m$:
\be
\tilde{\kappa } = \kappa  - 2\varphi ;\;\tilde{\tilde{\kappa}}= \kappa  - \varphi ;\; \kappa  = h + \nu _1(\hat{q}_0) m\,.
\ee
The system of equations for $m$, $\varphi$ which follows from(5.7) reads
\be
m =\tanh\beta\tilde{\kappa } = 4 \sinh 2 \beta \tilde{\tilde{\kappa}}
\,[b + \cosh 2 \beta \tilde{\tilde{\kappa}}]\, Z^{-1}_R(\tilde{\tilde{\kappa}})\,.
\ee
Evidently from the first equation of (5.15)we can obtain the expression for
$\varphi=\varphi(m)$ and we can reduce (5.15) to one equation for parameter $m$.

        Further we shall use also the following intracluster CF (we present
them only at $\kappa_i=\kappa$)
\bee
&&F^{(1)} = m = \tanh \beta  \tilde{\kappa };\qquad F^{(2)} = 1 - m^2\,, \nonumber\\
&&F_1(\tilde{\tilde{\kappa}}) = F^{(11)}_{12,R} = 2 {-d + \cosh 4\beta\tilde{\tilde{\kappa}}\over Z_R(\tilde{\tilde{\kappa}})} - m^2
\stackrel{T>T_c}{\rightarrow} {1-d\over 1+2a+4b+d} \,,\nonumber\\
&&F_2(\tilde{\tilde{\kappa}}) = F^{(11)}_{13,R} = 2 {+d-2a + \cosh 4\beta \tilde{\tilde{\kappa}}\over Z_R(\tilde{\tilde{\kappa}})} - m^2
\stackrel{T>T_c}{\rightarrow} {1-2a+d\over 1+2a+4b+d}\,, \nonumber\\
&&F_1 = F^{(11)}_{12,R} = F^{(11)}_{13,R} = F^{(11)}_{24,R} = F^{(11)}_{34,R}\,, \\
&&F_2 = F^{(11)}_{13,R} = F^{(11)}_{24,R}\,.\nonumber
\eee

\subsection{Pair correlation functions of the reference system.}

   The procedure for obtaining CFs of the reference system is similar to
mentioned above in the case of two-particle cluster. From (2.5) and (5.7) one
has
\beea
&&\ang{S_iS_j} = {\ F}^{(2)}_{ij} = F^{(2)}_i \tilde{\kappa }^{(1)}_{ij} \,,\\
&&\tilde{\kappa }^{(1)}_{ij} = {\delta \over \delta \kappa _j} \tilde{\kappa }_i = \delta _{ij} - \sum_{R'} \,_{R'}\varphi ^{(1)}_{ij} ; \,_{R'}\varphi ^{(1)}_{ij} = {\delta \over \delta \kappa _j} \,_{R'}\varphi _i\,, \nonumber\\
&&\tilde{\tilde{\kappa}}^{(1)}_{ij} = {\delta \over \delta \kappa _j}\tilde{\tilde{\kappa}}_i = \delta _{ij} - _{R_i}\varphi ^{(1)}_{ij} = \tilde{\kappa }^{(1)}_{ij} + \,_R\varphi _{ij}\,.
\eeea
For obtaining equation for $\tilde{\kappa}^{(1)}_{ij}$ we differentiate
(5.7b) with respect to $\kappa_f$. This leads to relation (the notations $R$,
$R_r$, $1$, $2_r$, $3_r$, $4_r$ correspond to figure 2)
\bee
&&F^{(2)}_1 \,_R\varphi ^{(1)}_{1j} + F^{(11)}_{12,R} \,_R\varphi ^{(1)}_{2j} + F^{(11)}_{13,R} \,_R\varphi ^{(1)}_{3j} + F^{(11)}_{14,R} \,_R\varphi ^{(1)}_{4j} = \nonumber\\
&&= - F^{(11)}_{12,R}\; \tilde{\kappa }^{(1)}_{2j} - F^{(11)}_{13,R}\; \tilde{\kappa }^{(1)}_{3j} - F^{(11)}_{14,R}\; \tilde{\kappa }^{(1)}_{4j}\,.
\eee
After cyclic permutation of the indices 1, 2, 3, 4, in (5.18) we shall find another
three relations of (5.18) type. One can write  for
cluster $R_1$ four equation of the same type. Let us present these equations for cluster $R_r$ ($r=0,1$) in
the matrix form
\bee
&&[ \hat{F}^{(2)}_{R_r} + \hat{F}^{(11)}_{R_r} ]\, _{R_r}\varphi ^{(1)}_{(r),j} = - \hat{F}^{(11)}_{R_r} \tilde{\kappa }^{(1)}_{(r),j}\,, \\
&&\hat{F}^{(2)} =\{F^{(2)}_f \delta_{ff'} \};\qquad \hat{F}^{(11)}_R =\{F^{(11)}_{\myhbox{ff'},R} (1 - \delta _{\myhbox{ff'}})\}\,, \nonumber\\
&&_{R_r}\hat{\varphi }^{(1)}_{(r),j}=\{_{R_r}\varphi ^{(1)}_{\myhbox{ij}}\} ;\qquad\hat{\tilde{\kappa}}^{(1)}_{(r),j} = \{\tilde{\kappa} ^{(1)}_{\myhbox{ij}}\}\,.
\eee
Here $F^{(2)}$, $F^{(11)}$ are the square matrices. They are formed by
indices $f,f'=1_r,2_r,3_r,4_r$. The columnar matrices $_{R_r}\varphi^{(1)}$,
$\hat{\tilde{\kappa}}^{(1)}_j$ are formed by index $j=1, 2, 3, 4$.

        In the following we consider only uniform case. Then on the basis
of two matrix equations (5.19) (writen at $r=0$ and $r=1$) and after some
transformations we obtain the matrix equation
\be
\pmatrix{-\tilde{\kappa }^{(1)}_{1j}+\delta _{1j}\cr\varphi ^{(1)}_{2j}+\varphi ^{(1)}_{2_1j}\cr\varphi ^{(1)}_{3j}+\varphi ^{(1)}_{3_1j}\cr\varphi ^{(1)}_{4j}+\varphi ^{(1)}_{4_1j}} =
- A \pmatrix{2 \tilde{\kappa }^{(1)}_{1j}\cr\tilde{\kappa }^{(1)}_{2j} + \tilde{\kappa }^{(1)}_{2_1j}\cr\tilde{\kappa }^{(1)}_{3j} + \tilde{\kappa }^{(1)}_{3_1j}\cr\tilde{\kappa }^{(1)}_{4j} + \tilde{\kappa }^{(1)}_{4_1j}}\,,
\ee
where
\beea
&&\hat{A} = \left[\hat{F}^{(2)}+\hat{F}^{(11}\right] \hat{F}^{(11)} \,,\\
&&\hat{F}^{(2)} =\{F^{(2)} \delta _{\myhbox{ff'}}\};\;
\hat{F}^{(11)} = \pmatrix{0&F_1&F_2&F_1\cr F_1&0&F_1&F_2\cr F_2&F_1&0&F_1\cr F_1&F_2&F_1&0}\,.
\eeea
It is necessary to obtain the closed equation for $\tilde{\kappa}^{(1)}_{ij}$
from (5.21). Therefore we find the first element at the left hand side of
(5.21)
\beea
&&- \tilde{\kappa }^{(1)}_{1j} + \delta _{1j} = - \sum_{\begin{array}{c}r\in R,R_1\\r\ne1\end{array}  }
{\overline A}_{1r} \tilde{\kappa }^{(1)}_{rj} - 2 {\overline A}_0 \tilde{\kappa }^{(1)}_{ij} \,,\\
&&r\ne1:\;{\overline A}_{1r} = {\overline A}_{1r_1} = A_{1r};\qquad {\overline A}_0 = A_{11}\,.
\eeea
Let us introduce matrices $M^{-1}$ and $\overline{M}^{-1}$. They are connected
one with other as $\bar{A}$ and $\bar{A}^{-1}$
\bee
&&M = (1-A)^{-1} = 1 + (F^{(2)})^{-1} F^{(11)} ;\qquad A = 1 - M^{-1}\,, \nonumber\\
&&(\overline{M}^{-1})_{1r} = (\overline{M}^{-1})_{1r_1} = (M^{-1})_{1r};\; (\overline{M}^{-1})_0 = (\overline{M}^{-1})_{ii}\,.
\eee
Then the equation (5.23) for $\tilde{\kappa}^{(1)}_{ij}$ takes the form
\be
\sum_{r\in R_1R_2}\{\delta _{1r} [1-2(\overline{M}^{-1})_0]-(\overline{M}^{-1})_{1r} \}\, \kappa ^{(1)}_{rj} = - \delta _{1j}\,.
\ee
Taking into account (5.17) we write the equation for the reference CF
$\hat{F}^{(2)}$
\be
\sum_{r\in R_1R_2}\{\delta _{1r} [ {1\over F^{(2)}} - 2 P_0 ] - P_{1r}\}F^{(2)}_{rj} = - \delta _{1j}\,.
\ee
Here the matrix $F$ is defined as
\be
\hat{P} = M^{-1}(F^{(2)})^{-1}\qquad \hat{P}^{-1} = F^{(2)} + F^{(11)}\,.
\ee
Let us return to complex notations of bonds $i=(\set{n}_i,\alpha_i)$ for relation
(5.26). We shall express the basic vectors of the Bravais lattice cell
$\set{a}_1$, $\set{a}_2$, $\set{a}_3$ in terms of vectors of the West's cell
$\set{e}_a$
\bee
&&{\bf a}_1 = {1\over 2} \left[-{\bf e}_1+{\bf e}_2+{\bf e}_3\right];\; {\bf a}_2 = {1\over 2} \left[{\bf e}_1+{\bf e}_2+{\bf e}_3\right];\; {\bf a}_3 = {\bf e}_2 \,,\nonumber\\
&&{\bf b}_1 = 2\pi  \left[-{\bf e}_1+{\bf e}_3\right];\; {\bf b}_2 = 2\pi  \left[{\bf e}_1+{\bf e}_3\right];\; {\bf b}_3 = 2\pi  \left[{\bf e}_2-{\bf e}_3\right] \,.
\eee
Here $\set{b}_a$ are the basic vectors of reciprocal lattice. Then after Fourier
transformation over $\set{n}$ the relation (5.26) takes the form
\bee
&&\sum_{\alpha _1}\{\delta _{\alpha \alpha _1} [ {1\over 1-m^2} - 2 P_0 ] - P_{\alpha \alpha _1}(\set{q})\}F^{(2)}_{\alpha _1\alpha '}(\set{q}) = - \delta _{\alpha \alpha '}\,, \nonumber\\
&&\{(1 - m^2)^{-1} - \hat{P}(\set{q})\}F^{(2)}(\set{q}) = - I\,.
\eee
Here the matrix $P(\set{q})$ has the form
\bee
&&P(\set{q})=\pmatrix{2P_0&P_3(\set{q})&P_1(\set{q})&P_2(\set{q})\cr
P_3(\set{q})&2P_0&P_5(\set{q})&P_6(\set{q})\cr
P_1(\set{q})&P_5(\set{q})&2P_0&P_4(\set{q})\cr
P_2(\set{q})&P_6(\set{q})&P_4(\set{q})&2P_0}\,, \\
&&P_1(\set{q}) = 2P_2 \cos  {q_x\over 2} ;\, P_2(\set{q}) = 2P_1 \cos  {1\over 4} (q_x - q_y + q_z)\,, \nonumber\\
&&P_3(\set{q}) = 2P_1 \cos  {1\over 4} (q_x + q_y + q_z);\, P_4(\set{q}) = 2P_1 \cos  {1\over 4} (q_x + q_y - q_z)\,, \nonumber\\
&&P_5(\set{q}) = 2P_1 \cos  {1\over 4} (q_x - q_y - q_z);\, P_6(\set{q}) = 2P_2 \cos  {1\over 4} q_y \,.
\eee
The quantities $P_0$, $P_1$, $P_2$ are expressed in the terms of functions $F^{(2)}$,
$F_1$, $F_2$ with the aid of the relation
\be
{1\over 2} P(\set{q} = 0) = P = \left[F^{(2)}+F^{(11)}\right]^{-1}\,.
\ee
Here $q_x$, $q_y$, $q_z$ are projections of dimensionless wave-vector
$\set{q}$ on the vectors $\set{e}_a$ and in the first Brillouin zone
$q_a\in[-\pi,\pi]$. From (5.32) we find
\beea
&&P_0 = {\cal D}^{-1}U_0 ; P_1 = {\cal D}^{-1}U_1 ; P_2 = {\cal D}^{-1}U_2 \,,\\
&&P_0 + 2P_1 + P_2 = (F^{(2)} + 2F_1 + F_2)^{-1} = {1\over 8} {Z_R\over c_4+bc_2-0.5 Z_R m^2}\,, \nonumber\\
&&P_0 - P_2 = (F^{(2)} - F_2)^{-1} = {1\over 8} {Z_R\over b c_2 + a}\,, \\
&&{\cal D} = {32\over Z_R} ( bc_2 + a) (bc_2 + d) (c_4 + bc_2 - 0.5 Z_r m^2)\,, \nonumber\\
&&U_0 = (3bc_2 + a + 2ab + 2d) + (4b + 3d) bc_2 +\myhbox{ ad }- \nonumber\\
&&- 0.5 Z_R m^2 (2c_4 + 2bc_2 + d) - {1\over 8} Z^2_R m^2 \,,\nonumber\\
&&U_1 = - (c_4 - d) (bc_2 + a) + 0.5 Z_R m^2 (bc_2 + a)\,, \\
&&U_2 = (-bc_2 + a - 2d) c_4 + (2a - d) bc_2 +\myhbox{ ad }- \nonumber\\
&&- 0.5 Z_R m^2 (- bc_2 + a - 2d) - {1\over 8} Z^2_R m^4 \,,\nonumber\\
&&c_4 = \cosh 4 \beta \tilde{\tilde{\kappa}};\qquad c_2 = \cosh 2\beta \tilde{\tilde{\kappa}}\,.\nonumber
\eeea
Let us note that relations of (5.29) at $T>T_c$ were obtained firstly in~[34]. The method developed in this work is acceptable only in the case $T>T_c$.
Moreover in~[34] the Sleter model corresponding to the Hamiltonian (5.1) at
$\Phi=0$, was considered. For the mentioned above cases the matrix (5.30)
coincides with corresponding matrix of~[34].

     The components of susceptibility tenzor (see next subsection ) will
contain only certain eigenvalues of matrix $P(\bf q)$. Therefore we shall consider
the procedure of diagonalizing of equation (5.29). Let us introduce notation
for eigenfunctions $U_{f\mu}(\bf q)$ and eigenvalues
 $\epsilon_\mu(\bf q)$ of matrix $\hat{P}(\set{q})-2P_0\hat{I}$.
After unitary transformation we obtain
\bee
&&\hat{\tilde{\F}}^{(2)}(\set{q}) = \hat{U}^+(\set{q}) \hat{\F}^{(2)}(\set{q}) \hat{U}(\set{q});\qquad \hat{\F}^{(2)}(\set{q}) = \hat{U}(\set{q})\hat{\tilde{\F}}^{(2)}(\set{q}) \hat{U}^+(\set{q})\,, \nonumber\\
&&\F^{(2)}_{\myhbox{ff'}}(\set{q}) = \sum_{\mu } U_{f\mu }(\set{q}) \tilde{\F}^{(2)}_{\mu }(\set{q}) U_{f'\mu }(\set{q})\,, \nonumber\\
&&\tilde{\F}^{(2)}_{\mu }(\set{q}) = \left[\epsilon _{\mu }(\set{q})+2P_0-(1-m^2)^{-1}\right]^{-1}\,.
\eee
Here quantities $\epsilon_{\mu}(\bf q)$ are to be found from the equation
\bee
&&\epsilon ^4 - [ \sum^6_{f=1} P^2_f(\set{q}) ] \epsilon ^2 - 2 \left[ P_1(\set{q}) \left(P_2(\set{q})P_4(\set{q})+P_3(\set{q})P_5(\set{q})\right) +\right. \nonumber\\
&&\left.+ P_6(\set{q}) \left(P_2(\set{q})P_3(\set{q})+P_4(\set{q})P_5(\set{q})\right)\right]\epsilon  + P^2_1(\set{q}) P^2_6(\set{q}) + \nonumber\\
&&+ P^2_2(\set{q}) P^2_5(\set{q}) + P^2_3(\set{q}) P^2_4(\set{q}) - 2P_2(\set{q}) P_3(\set{q}) P_4(\set{q}) P_5(\set{q}) - \nonumber\\
&&- 2P_1(\set{q}) P_6(\set{q}) \left[P_2(\set{q})P_5(\set{q})+P_3(\set{q})P_4(\set{q})\right] = 0 \,.
\eee
We consider such values $\set{q}$ for which the problem of eigenfunctions and eigenvalues
due to the symmetry of $P(\set{q})$ may lead to solution the equations of second
order. For the values $\set{q}=(q,q,q_z)$ we find
\beea
&&2\epsilon _{1,2}(\set{q}) = P_3(\set{q}) + P_4(\set{q}) \pm  \sqrt{{\cal D}_{12}^+ (\set{q})}\,. \nonumber\\
&&2\epsilon _{3,4}(\set{q}) = - P_3(\set{q}) - P_4(\set{q}) \pm  \sqrt{{\cal D}_{12}^- (\set{q})} \,,\\
&&{\cal D}^{\pm }_{12}(\set{q}) = \left[P_3(\set{q})-P_4(\set{q})\right]^2 + 4 \left[P_1(\set{q})\pm P_2(\set{q})\right]^2\,; \nonumber\\
&&U(\set{q})=\pmatrix{\varphi_{11}^+(\set{q})&\varphi_{12}^+(\set{q})&\varphi_{11}^-(\set{q})&\varphi_{12}^-(\set{q})\cr
\varphi_{11}^+(\set{q})&\varphi_{12}^+(\set{q})&-\varphi_{11}^-(\set{q})&-\varphi_{12}^-(\set{q})\cr
\varphi_{12}^+(\set{q})&-\varphi_{11}^+(\set{q})&\varphi_{12}^-(\set{q})&-\varphi_{11}^-(\set{q})\cr
\varphi_{12}^+(\set{q})&-\varphi_{11}^+(\set{q})&-\varphi_{12}^-(\set{q})&\varphi_{11}^-(\set{q})}\,; \\
&&\varphi ^{\delta }_{11}(\set{q}) = - \varphi ^{\delta }_{22}(\set{q}) = {1\over \sqrt{2}} \left[\matrix{1+\delta \,r_{\delta }(\set{q})}\right]^{1/2}\,, \nonumber\\
&&\varphi ^{\delta }_{12}(\set{q}) = {[1-\delta \,r_{\delta }(\set{q})]^{1/2}\over \sqrt{2}}\,, \\
&&r_{\delta }(\set{q}) = {1\over 2} R_{\delta }(\set{q}) \left[\matrix{1+{1\over 4}\,R^2_{\delta }(\set{q})}\right]^{1/2};
\; R_{\delta }(\set{q}) = {P_3(\set{q}) - P_4(\set{q})\over P_1(\set{q})+\delta  P_2(\set{q})}\,.\nonumber
\eeea
For the values $\set{q}=(q_x,q_y,0)$ we obtain
\beea
&&2\epsilon _{1,3}(\set{q}) = P_1(\set{q}) + P_6(\set{q}) \pm  \sqrt{{\cal D}^{+}_{32} (\set{q})}\,, \nonumber\\
&&2\epsilon _{2,4}(\set{q}) = - P_1(\set{q}) - P_6(\set{q}) \pm  \sqrt{{\cal D}^{-}_{32} (\set{q})}\,, \\
&&{\cal D}^{\pm }_{32}(\set{q}) = \left[\matrix{P_1(\set{q})&-&P_6(\set{q})}\right]^2 + 4 \left[\matrix{P_3(\set{q})&\pm &P_2(\set{q})}\right]^2 \,;\nonumber\\
&&U(\set{q})=\pmatrix{\varphi_{11}^+(\set{q})&\varphi_{12}^+(\set{q})&\varphi_{11}^-(\set{q})&\varphi_{12}^-(\set{q})\cr
\varphi_{12}^+(\set{q})&\varphi_{11}^+(\set{q})&\varphi_{12}^-(\set{q})&-\varphi_{11}^-(\set{q})\cr
\varphi_{11}^+(\set{q})&\varphi_{12}^+(\set{q})&-\varphi_{11}^-(\set{q})&-\varphi_{12}^-(\set{q})\cr
\varphi_{12}^+(\set{q})&-\varphi_{11}^+(\set{q})&-\varphi_{12}^-(\set{q})&\varphi_{11}^-(\set{q})}\,, \\
&&\varphi ^{\delta }_{11}(\set{q}) = - \varphi ^{\delta }_{22}(\set{q}) = {[1+\delta \,r_{\delta }(\set{q})]^{1/2}\over \sqrt{2}}\,,  \nonumber\\
&&\varphi ^{\delta }_{12}(\set{q}) = {[1-\delta \,  r_{\delta }(\set{q})]^{1/2}\over \sqrt{2}}\,, \\
&&r_{\delta }(\set{q}) = {R_{\delta }(\set{q})\over 2} \left[1+{1\over 4}R^2_{\delta }(\set{q})\right]^{1/2};
\;R_{\delta }(\set{q}) = {P_1(\set{q}) - P_6(\set{q})\over P_3(\set{q})+\delta  P_2(\set{q})}\,.\nonumber
\eeea
Let us note that in ~[34] the eigenvalues were obtained only for
vector $\set{q}=(q,q,q_z)$ and the long-range interaction was not considered.

\subsection{Susceptibility of $KD_2PO_4$ system.}

   The tensor of isotermical susceptibility of $KD_2PO_4$ system can be expressed
in terms of the pair CF  $^L\F^{(2)}_{\alpha\beta}(\set{q})$ taking into
consideration long-range interaction in MFA in the following way

\beea
&&\chi _{ab}(\set{q}) = \sum_{\alpha ,\beta } d_{\alpha a} d_{\beta b} \,^L\F^{(2)}_{\alpha \beta }(\set{q})\,, \nonumber\\
&&^L\hat{\F}^{(2)}(\set{q}) =\left\{\left[\hat\F^{(2)}(\bf q)\right]^{-1} - \beta  \hat{J}(\set{q})\right\}^{-1}\,, \\
&&d_{\alpha z} = d_c;\qquad d_x = d_{1x} = - d_{3x};\qquad d_{2x} = d_{4x} = 0\,, \nonumber\\
&&d_y = d_{2y} = - d_{4y};\qquad d_{1y} = - d_{3y} = 0\,.
\eeea
Here $d_{\alpha a}$ is an effective dipolar moment of hydrogen bond.
It contains the contribution of ionic subsystem via spin-phonon interaction.
We write the longitiudinal and transversal components of tenzor
$\chi_{ab}(\set{q})$ in terms of eigenfunctions $W_{\alpha\mu}(\set{q})$
and eigenvalues $\tilde\F^{(2)}_\mu(\set{q})$ of the matrix
$^L\F^{(2)}_{\alpha\beta}(\set{q})$.
\bee
&&\chi _{zz}(\set{q}) = d^2_c \sum_{\mu } \tilde\F^{(2)}_{\mu }(\set{q}) f^2_{\mu }(\set{q});\; f_{\mu }(\set{q}) = \sum_f W_{f\mu }(\set{q}), \nonumber\\
&&\chi _{xx}(\set{q}) = d^2_x \sum_{\mu } \tilde\F^{(2)}_{\mu }(\set{q}) \left[W_{1\mu }(\set{q})-W_{3\mu }(\set{q})\right]^2 \,, \\
&&\chi _{yy}(\set{q}) = d^2_y \sum_{\mu } \tilde\F^{(2)}_{\mu }(\set{q}) \left[W_{2\mu }(\set{q})-W_{4\mu }(\set{q})\right]^2 \,. \nonumber
\eee
The symmetry of matrix $\hat\F^{(2)}(\set{q})$ which is related with the symmetry of matrix
$\hat{P}(\set{q})$ at arbitrary value of $\set{q}$ does not coincide with the symmetry of matrix
$J_{\alpha\beta}(\set{q})$. Here we consider only such values of $\set{q}$ when $P(\set{q})$ and $I(\set{q})$ has
equal eigenfunctions. Then on the basis (5.38a) and (5.34) one can write
\bee
&&\tilde\F^{(2)}_{\mu }(\set{q}) = \left[-(1-m^2)^{-1}+2P_0+\epsilon _{\mu }(\set{q})-\nu _{\mu }(\set{q})\right]^{-1} \,, \nonumber\\
&&\epsilon _{\mu }(\set{q}) =\sum_{\alpha ,\alpha '} W_{\alpha \mu }(\set{q}) P_{\alpha \alpha '}(\set{q}) W_{\alpha '\mu }(\set{q}) - 2P_0 \, ,\nonumber\\
&&\nu _{\mu }(\set{q}) =\sum_{\alpha ,\alpha '} W_{\alpha \mu }(\set{q}) J_{\alpha \alpha '}(\set{q}) W_{\alpha '\mu }(\set{q}).
\eee
We present here several cases of application of relation (5.40). At $\set{q}\rightarrow 0$ we
use the expressions for $J_{\alpha\beta}(\set{q})$ from~[59]

1. $\set{q}=\hat{q}_z=(0,0,q_z)$
\beea
&&\hat{J}(\hat{q}_z) =\pmatrix
{J_{11}(\hat{q}_z)&J_{12}(\hat{q}_z)&J_{13}(\hat{q}_z)&J_{12}(\hat{q}_z)\cr
 J_{12}(\hat{q}_z)&J_{11}(\hat{q}_z)&J_{12}(\hat{q}_z)&J_{13}(\hat{q}_z)\cr
 J_{13}(\hat{q}_z)&J_{12}(\hat{q}_z)&J_{11}(\hat{q}_z)&J_{12}(\hat{q}_z)\cr
 J_{12}(\hat{q}_z)&J_{13}(\hat{q}_z)&J_{12}(\hat{q}_z)&J_{11}(\hat{q}_z)} \,,\\
&&J_{11}(\hat{q}_z) \stackrel{\hat{q}_z\rightarrow 0}{\longrightarrow}J_{11} + a_z;\qquad J_{12}(\hat{q}_z)
\stackrel{q_z\rightarrow 0}{\longrightarrow}J_{12}\,, \nonumber\\
&&J_{13}(\hat{q}_z) \stackrel{q_z\rightarrow 0}{\longrightarrow}J_{13} - a_z \,,\\
&&\hat{W}(\hat{q}_z) \equiv  \hat{U}(\hat{q}_z)
= {1\over 2}\pmatrix{1&1&1&1\cr1&1&-1&-1\cr1&-1&1&-1\cr1&-1&-1&1} \,,\\
&&\nu _{1,3}(\hat{q}_z)
= J_{11}(\hat{q}_z) + J_{13}(\hat{q}_z) \pm  2J_{12}(\hat{q}_z)
\stackrel{q_z\rightarrow 0}{\longrightarrow}J_{11} + J_{13} \pm  2J_{12} \,,\nonumber\\
&&\nu _{2,4}(\hat{q}_z) = J_{11}(\hat{q}_z) - J_{13}(\hat{q}_z)
\stackrel{q_z\rightarrow 0}{\longrightarrow} J_{11} - J_{13} \pm  2a_z \,,\\
&&\epsilon _{1,2}(\hat{q}_z) = 2 [ P_2 \pm  P_1 \cos  {q_z\over 4}]; \;\epsilon _{3,4}(\hat{q}_z) = - 2P_2 \,,\nonumber\\
&&{\chi _{zz}(\hat{q}_z)\over (2d_z)^2} =\{2(P_0 + 2P_1 + P_2) - 8P_1 \sin ^2 {q_z\over 8} - (1-m^2)^{-1} - \nu _1(q_z)\}^{-1} \,,\nonumber\\
&&{\chi _{aa}(\hat{q}_z)\over 2(2d_a)^2} =\{2(P_0 - P_2) - (1-m^2)^{-1} - \nu _2(q_z)\}^{-1};\, a = x,y\,.
\eeea

2. $\set{q}=\hat{q}=(\epsilon,\epsilon,0)$, $\epsilon\rightarrow 0$
\beea
&&J(\hat{q}) =\pmatrix
{J_{11}(\hat{q})&J_{12}(\hat{q})&J_{13}(\hat{q})&J_{14}(\hat{q})\cr
 J_{12}(\hat{q})&J_{11}(\hat{q})&J_{14}(\hat{q})&J_{13}(\hat{q})\cr
 J_{13}(\hat{q})&J_{14}(\hat{q})&J_{11}(\hat{q})&J_{12}(\hat{q})\cr
 J_{14}(\hat{q})&J_{13}(\hat{q})&J_{12}(\hat{q})&J_{11}(\hat{q})} \,,\\
&&J_{11}(\hat{q}) = J_{11} + a_{xy};\qquad J_{12}(\hat{q}) = J_{12}+a_{xy} \,,\nonumber\\
&&J_{14}(\hat{q}) = J_{12} - a_{xy};\qquad J_{13}(\hat{q}) = J_{13}-a_{xy} \,,\\
&&\hat{W}(\hat{q}) = \hat{U}(\hat{q})
= {1\over 2}\pmatrix{1&1&1&1\cr1&1&-1&-1\cr1&-1&1&-1\cr1&-1&-1&1} \,,\\
&&\nu _{1,3}(\hat{q}) = J_{11}(\hat{q}) + J_{13}(\hat{q}) \pm  [J_{12}(\hat{q})+J_{14}(\hat{q})]
= J_{11} + J_{13} \pm  2J_{12} \nonumber\,,\\
&&\nu _{2,4}(\hat{q}) = J_{11}(\hat{q}) - J_{13}(\hat{q}) \pm  [J_{12}(\hat{q})-J_{14}(\hat{q})] \nonumber\,,\\
&&\nu _2(\hat{q}) = J_{11} - J_{13}+4a_{xy};\qquad \nu _4(\hat{q}) = J_{11} - J_{13} \,,\\
&&\epsilon _{1,3}(\hat{q}) = \pm4P_1+2P_2;\qquad \epsilon _{2,4}(\hat{q}) = - 2P_2 \nonumber\,,\\
&&{\chi _{zz}(\hat{q})\over (2d_z)^2} =\{2(P_0 + 2P_1 + P_2) - (1-m^2)^{-1} - \nu _1(\hat{q})\}^{-1} \nonumber\,,\\
&&{\chi _{aa}(\hat{q})\over (2d_a)^2} =\sum_{\mu=2,4}\{2(P_0 - P_2) - (1-m^2)^{-1} - \nu_\mu(\hat{q})\}^{-1}\,.
\eeea

3. $\set{q}=\hat{a}$, $\hat{x}=(\epsilon,0,0)$, $\hat{y}=(0,\epsilon,0)$,
$\epsilon\rightarrow 0$
\beea
&&J(\hat{a}) =\pmatrix
{J_{11}(\hat{a})&J_{12}(\hat{a})&J_{13}(\hat{a})&J_{12}(\hat{a})\cr
 J_{12}(\hat{a})&J_{22}(\hat{a})&J_{12}(\hat{a})&J_{24}(\hat{a})\cr
 J_{13}(\hat{a})&J_{12}(\hat{a})&J_{11}(\hat{a})&J_{12}(\hat{a})\cr
 J_{12}(\hat{a})&J_{24}(\hat{a})&J_{12}(\hat{a})&J_{22}(\hat{a})} \,,\\
&&J_{11}(\hat{a}) = J_{11} + b(\hat{a});\qquad J_{22}(\hat{a}) = J_{11}+a(\hat{a}) \nonumber\,,\\
&&J_{13}(\hat{a}) = J_{13} - b(\hat{a});\qquad J_{24}(\hat{a}) = J_{13}-a(\hat{a}) \,,\\
&&J_{12}(\hat{a}) = J_{12};\qquad b(\hat{a}) = a(\hat{a}) \nonumber\,,\\
&&\hat{W}(\hat{a}) = \hat{U}(\hat{a})
= {1\over 2}\pmatrix{1&1&\sqrt{2}&0\cr1&-1&0&\sqrt{2}\cr1&1&-\sqrt{2}&0\cr1&-1&0&-\sqrt{2}} \,,\\
&&\nu _{1,3}(\hat{q}) = J_{11}(\hat{a}) + J_{13}(\hat{a}) \pm  J_{12}(\hat{a}) = J_{11} + J_{13} \pm  2J_{12} \nonumber\,,\\
&&\nu _2(\hat{q}) = J_{11}(\hat{a}) - J_{13}(\hat{a}) =J_{11}-J_{13}+2b(\hat{a}) \nonumber\,,\\
&&\nu _4(\hat{q}) = J_{22}(\hat{a}) - J_{24}(\hat{a}) = J_{11} - J_{13}+2a(\hat{a}) \,,\\
&&\epsilon _{1,2}(\hat{q}) = \pm 4P_1+2P_2;\qquad \epsilon _{3,4}(\hat{q}) = - 2P_2 \nonumber\,,\\
&&{\chi _{zz}(\hat{q})\over 2(2d_z)^2} =\{2(P_0 + 2P_1 + P_2) - (1-m^2)^{-1} - \nu _1(\hat{q})\}^{-1} \,,\\
&&{\chi _{aa}(\hat{q})\over 2(2d_a)^2} =\{2(P_0 - P_2) - (1-m^2)^{-1} - \nu_\mu(\hat{q})\}^{-1};\;\mu=\left\{\begin{array}{lc}2,&a=x\\4,&a=y\end{array}\right.\nonumber \,.
\eeea
More detailed investigation of $\vec{q}$-dependence for susceptibility of
$KD_2PO_4$ system will be carried out in the separate paper.

\section{Conclusion}

	Thus in the present work we have suggested the cluster method for
computation of the correlation functions of the Ising-type systems with
short-range interactions and with arbitrary value of spin. The method is
based on the calculation of generating function (that is the logarithm of
partition function for a system in nonuniform external field). Here we
restricted ourselves only to the first order of cluster approximation for
generating function. The investigation of some models ($S^z=\pm 1$, $S^z=-2,
0, +2$) has been carried out within two-particle cluster approximation. The
static susceptibility of $KD_2PO_4$  crystal has been
obtained within four-particle cluster approximation.

	The obtained here form of equations for CFs is similar to the form
which one can obtain within random phase approximation (molecular field
approximation for generating function). Moreover in the case of Bravais
lattice the $\set{q}$-dependence of CFs is similar to one obtained within
random-phase approximation. Nevertheless within cluster approximation we
obtain the more precise expressions for the temperature depending coefficients.
In the case of tree-like lattices, in particular, for 1D lattices, the
cluster approximation gives the exact expression for all static
characteristics of the Ising-type systems. The more detailed study
including numerical calculations of expressions obtained in the present work
will be given elsewhere.

\end{document}